\documentclass[english,aps,prb, superscriptaddress,floatfix, citeautoscript, notitlepage]{revtex4-1}

\pdfoutput = 1
\usepackage{blindtext}
\usepackage{siunitx}
\usepackage{graphicx} 
\usepackage{amsmath}
\usepackage{amssymb}
\usepackage{ bbold }
\usepackage{placeins} 
\usepackage{braket}
\usepackage{color}
\usepackage[table]{xcolor}

\RequirePackage[T1]{fontenc}
\RequirePackage[utf8]{inputenc}
\RequirePackage{lmodern}

\usepackage[colorlinks=true,allcolors=blue]{hyperref}
\usepackage{units}
\usepackage[toc,page]{appendix}

\newcommand{\rfig}[1]{Fig.\,\ref{#1}}
\newcommand{\rapp}[1]{\ref{#1}}
\newcommand{\req}[1]{Eq.\,(\ref{#1})}
\newcommand{\rsec}[1]{Sec.\,\ref{#1}}      
 
\newcommand{\rtab}[1]{Tab.\,\ref{#1}}  
\newcommand{\rref}[1]{Ref.\,\cite{#1}}

\begin{document}

\title{Full configuration interaction simulations of exchange-coupled donors in silicon using multi-valley effective mass theory}

\author{Benjamin Joecker}
\affiliation{Centre for Quantum Computation and Communication Technology, School of Electrical Engineering \& Telecommunications, UNSW, Sydney, NSW, 2052, Australia}
\author{Andrew D. Baczewski}
\affiliation{Center for Computing Research, Sandia National Laboratories, Albuquerque, NM, 87123, USA}
\author{John K. Gamble}
\affiliation{Center for Computing Research, Sandia National Laboratories, Albuquerque, NM, 87123, USA}
\affiliation{Microsoft Quantum, 1 Microsoft Way, Redmond, WA, 98052, USA}
\author{Jarryd J. Pla}
\affiliation{Centre for Quantum Computation and Communication Technology, School of Electrical Engineering \& Telecommunications, UNSW, Sydney, NSW, 2052, Australia}
\author{Andr{\'e} Saraiva}
\affiliation{Centre for Quantum Computation and Communication Technology, School of Electrical Engineering \& Telecommunications, UNSW, Sydney, NSW, 2052, Australia}
\author{Andrea Morello}
\affiliation{Centre for Quantum Computation and Communication Technology, School of Electrical Engineering \& Telecommunications, UNSW, Sydney, NSW, 2052, Australia}

\begin{abstract}
		Donor spin in silicon have achieved record values of coherence times and single-qubit gate fidelities. The next stage of development involves demonstrating high-fidelity two-qubit logic gates, where the most natural coupling is the exchange interaction. To aid the efficient design of scalable donor-based quantum processors, we model the two-electron wave function using a full configuration interaction method within a multi-valley effective mass theory. We exploit the high computational efficiency of our code to investigate the exchange interaction, valley population, and electron densities for two phosphorus donors in a wide range of lattice positions, orientations, and as a function of applied electric fields. The outcomes are visualized with interactive images where donor positions can be swept while watching the valley and orbital components evolve accordingly. Our results provide a physically intuitive and quantitatively accurate understanding of the placement and tuning criteria necessary to achieve high-fidelity two-qubit gates with donors in silicon.
\end{abstract}
	
\maketitle

\section{Introduction}
Donor spins in silicon are prominent candidates for the construction of a large-scale quantum computer.
The spin-$1/2$ degree of freedom of an electron bound to a donor atom is a natural qubit. Isolated in the environment of isotopically purified $^{28}$Si, it has demonstrated exceptional coherence times\cite{muhonen2014storing}.
High fidelity initialization, readout \cite{morello2010single}, and single-qubit operations\cite{pla2012single,muhonen2015quantifying,dehollain2016optimization} are well established.
Several proposals for multi-qubit gates rely on the exchange interaction, using either strong tunable exchange to perform SWAP oscillations \cite{kane1998silicon,hollenberg2006two,hill2007robust}, or weak exchange and microwave pulses to perform a CROT gate \cite{hill2007robust,kalra2014robust}.
Experimental work by He \textit{et al.}\cite{he2019two} demonstrated fast SWAP oscillations in the strong exchange regime, using two donor clusters  fabricated with a scanning tunnel microscope.
Madzik \textit{et al.}\cite{mkadzik2020conditional} showed an embryonic CROT gate as the coherent rotation of a target qubit, conditional an a specific state of the control, using a pair of ion-implanted $^{31}$P donors, in the weak exchange regime.

Moving from these proof-of-principle demonstrations to systematic achievement of high-fidelity two-qubit gates will be greatly assisted by having an accurate and efficient tool to simulate the electronic structure of coupled donors. Here, we achieve this goal by developing and applying a full configuration interaction implementation of effective mass theory that avoids unjustified approximations made in some earlier theories. Our code can be made computationally very efficient, which allows us to study detailed properties of coupled donors as a function of a very wide range of parameters. This also allows us to pedagogically unpack the microscopic physics that determines the results. We trust that the physically transparent nature of this work will help future researchers gaining better insights into the subtle physics of donor spin qubits.

Effective mass theory (EMT) was originally developed to find approximate solutions for the complex wave-function of a donor-bound electron in silicon\cite{kohn1955theory,luttinger1955motion} and has been continuously developed\cite{ning1971multivalley,pantelides1974theory,shindo1976effective,hui2013improved,saraiva2015theory}. It has been applied to donors near an interface\cite{calderon2009quantum,baena2012impact} and single-electron donor molecules\cite{hu2005charge,klymenko2014electronic,gamble2015multivalley,klymenko2017electronic,saraiva2015theory}. It is widely used to model physical properties, such as valley splitting\cite{saraiva2011intervalley,gamble2016valley}, tunnel coupling\cite{hu2005charge,gamble2015multivalley}, and both Stark\cite{friesen2005theory,debernardi2006computation} and hyperfine\cite{pica2014hyperfine} shifts.

Exchange-coupled donor molecules (Fig.~\ref{Fig:1}a) are many-electron systems and computationally very complex.
Their study requires the application of quantum chemical methods and approximations \cite{szabo2012modern,jensen2017introduction} to the multi-valley effective mass theory frame work.
Widely used is the Heitler-London approach\cite{koiller2001exchange,wellard2003electron,wellard2005donor,pica2014exchange}. Here, the two-electron wave functions are simply the symmetrized and anti-symmetrized products of the donor ground state orbitals.
Early work by Koiller \textit{et al.}\cite{koiller2001exchange} and Wellard \textit{et al.}\cite{wellard2003electron} approximated the donor potential by a simple Coulomb potential and ignored the valley-orbit coupling (VOC) terms.
Extending on this work, Wellard \textit{et al.}\cite{wellard2005donor} and Pica \textit{et al.}\cite{pica2014exchange} included VOC and an isotropic central cell correction (CCC) to correct for the short range limit where the dielectric constant becomes meaningless and the impurity deforms the charge density that the electron interacts with. However, they failed to precisely reproduce the excited states within the $1s$ subspace. In the Hartree-Fock (HF) method the many-electron wave function is constructed from an antisymmetrized product of optimized molecular orbitals. Wu \textit{et al.}\cite{wu2020excited} applied HF to an EMT implementation that includes a CCC and VOC. Their effective mass tensor is anisotropic, while their Gaussian basis is not.

To go beyond the HF, the Configuration Interaction (CI) method systematically adds more configurations, i.e. the symmetrized and antisymmetrized products of orbitals. 
The  Hund-Mulliken  approach  adds  ionic  configurations  to  the two-electron basis\cite{kettle2006molecular}.
Including all possible configurations is called Full CI and it is the optimal solution in a given valley-orbit basis\cite{szabo2012modern,jensen2017introduction,saraiva2015theory}. Gonzalez-Zalba \textit{et al.}\cite{gonzalez2014exchange} and Saraiva \textit{et al.}\cite{saraiva2015theory}
also implemented Full CI to model exchange-coupled donors.
However, they used a hydrogenic model, i.e. an isotropic basis and effective mass, and neglected VOC.
Additionally, they only applied a CCC to the donor ground state and assumed the excited states to be degenerate.

Our theory builds upon the multi-valley EMT implementation by Gamble \textit{et al.}\cite{gamble2015multivalley,gamble2016valley}.
They solved the coupled six-valley Shindo-Nara equations, including silicon’s full Bloch functions\cite{shindo1976effective}. 
They introduced a new tetrahedral CCC and went beyond a minimal basis set.
As a result, they were able to precisely reproduce the entire spectrum of valley-orbit split $1s$ states.
The efficient use of Gaussian type orbitals allowed them to study $\sim\SI{1.3e6}{}$ relative lattice placements of a single-electron two-phosphorus donor molecule.

Here, we add an efficient HF and Full CI solver to study exchange-coupled phosphorus donors. We introduce a efficient basis set of contracted Gaussian orbitals, to reduce computational costs. We perform a basis set analysis and find a fast and accurate choice. This allows us to perform exhaustive 3D iterations over all physically relevant relative donor placements for a given distance. We study the exchange interaction and valley configurations in detail for the [100], [110] and [111] crystal orientations and visualize the electron density. We find that a CROT gates are viable over a range of distances between \SI{10}{\nano\meter} and \SI{24}{\nano\meter} depending on the crystal direction. Along [100] and [110] the donor placement may straggle by \SI{5}{\nano\meter}. Finally, we study the electric tunability of exchange-coupled donors. Using a simple noise model, we find that high-quality SWAP gates are viable.

The paper is organized as follows. In \rsec{Sec:EMT}, we review the multi-valley effective mass theory as used by Gamble \textit{et al.}\cite{gamble2015multivalley,gamble2016valley}. \rsec{Sec:QuantumChemistry} introduces  the HF and CI methods, applied to donors in silicon. In \rsec{Sec:BasisOpti}, we discuss our choice of basis functions. In \rsec{Sec:ExchangeCoupledDonors}, we use our theory to calculate the key properties of exchange-coupled donors. \rsec{Sec:Comparison} compares our results to earlier work. \rsec{Sec:Tunability} discusses the exchange tunability with electric fields.
In \rsec{Sec:Conclusion}, we summarize our main conclusions.

\section{Effective Mass Theory}
\label{Sec:EMT}

\begin{figure}[htb!]
	\centering
	\includegraphics[width=0.5\columnwidth]{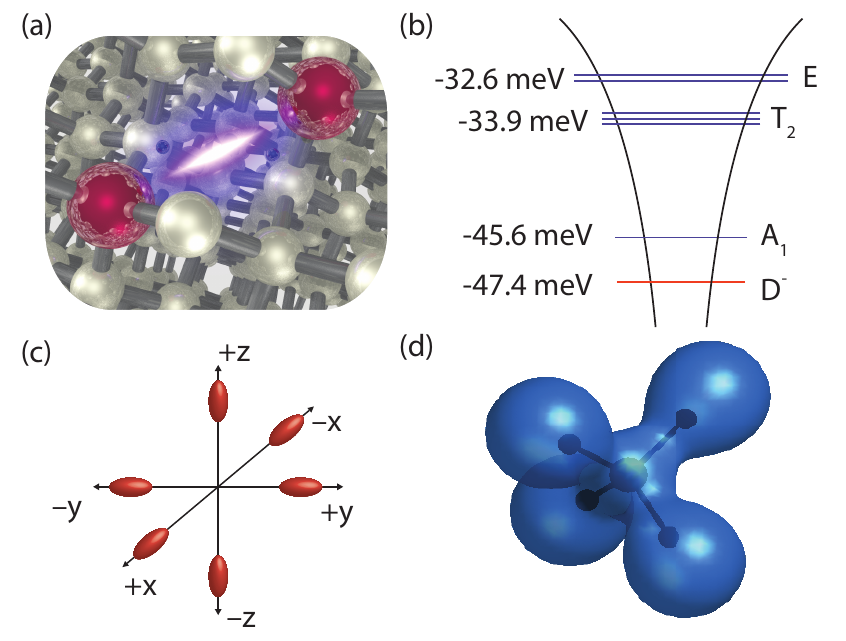}
	\caption{(a) Illustration of an exchange-coupled pair of phosphorus donors (red) in a lattice of silicon atoms (silver) (b) Experimentally measured energy levels of a negative\cite{narita1982uniaxial} (red) and neutral donor\cite{grimmeiss1982multivalley} (blue). The degeneracy of the six conduction band minima (`valleys') illustrated in panel (c) is lifted by the valley-orbit coupling. (d) Constant energy surfaces $E=\SI{-1}{\electronvolt}$ of the impurity potential with the tetrahedral central cell correction in \req{Eq:CC} with parameters given in \rtab{Tab:SmallBasis}.}
	\label{Fig:1}
\end{figure}
This section gives a brief review of the underlying effective mass theory established by Gamble \textit{et al.}\cite{gamble2015multivalley}.
The cornerstone of EMT is that low-energy conduction electron states $\psi(\vec{r})$ in silicon only have support around the six valley minima (\rfig{Fig:1}c):
\begin{align}
	\psi(\vec{r})&=\sum_{\mu}F^\mu(\vec{r})\varphi_\mu(\vec{r}).
	\label{Eq:EMTstate}
\end{align}
Here, we sum over the six valley minima at $\vec{k}_\mu$ with the corresponding Bloch functions $\varphi_\mu(\vec{r})=u_{\vec{k}_\mu}(\vec{r})e^{i\vec{k}_\mu\vec{r}}$.
The envelope functions $F^\mu(\vec{r})$ are subject to the multivalley EMT equations, a system of coupled Schr\"odinger-like equations:
\begin{align}
	[\mathbf{T}_\mu +U(\vec{r})]F^\mu(\vec{r})+\sum_\mu V^{VO}_{\mu\nu}(\vec{r})F^\nu(\vec{r}) = E F^\mu(\vec{r}).
	\label{Eq:EMTEqs}
\end{align}
The kinetic energy operator of the $\mu$th valley $\mathbf{T}_\mu$ gives a different effective mass to motions along ($m_{\parallel}$) and perpendicular to ($m_{\perp}$) $\vec{k}_\mu$, e.g. $\mathbf{T}_{+x}=-\frac{\hbar^2}{2m_{\parallel}}\frac{\partial^2}{\partial x^2}-\frac{\hbar^2}{2m_{\perp}}(\frac{\partial^2}{\partial y^2}+\frac{\partial^2}{\partial z^2})$.
$U(\vec{r})$ is the potential of the impurity.
The EMT equations are coupled via the valley-orbit potential 
\begin{align}
	V^{VO}_{\mu\nu}(\vec{r})=\varphi^\ast_\mu(\vec{r})\varphi_\nu(\vec{r})U(\vec{r}),
	\label{Eq:VOC}
\end{align}
which lifts the degeneracy of the valley states and results in the $A_1$, $T_2$ and E valley-orbit states depicted in \rfig{Fig:1}b.
The relative amplitudes of $F^\mu(\vec{r})$ for each of these states are governed by the the symmetry of the crystal, namely $T_d$.
We give the relative amplitudes in vector form in natural ordering ($+x$, $-x$, $+y$, ...),
\begin{align}
	& A_1 \quad\frac{1}{\sqrt{6}}(1,1,1,1,1,1), 
	\label{Eq:A1}\\
	& T_2 \quad
	\begin{cases}
		\frac{1}{\sqrt{2}}(1,-1,0,0,0,0)\\
		\frac{1}{\sqrt{2}}(0,0,1,-1,0,0)\\
		\frac{1}{\sqrt{2}}(0,0,0,0,1,-1),~\text{and}
	\end{cases}
	\label{Eq:T2}
	\\
	& E\; \quad
	\begin{cases}
		\frac{1}{2}(1,1,-1,-1,0,0)\\
		\frac{1}{2}(1,1,0,0,-1,-1).
	\end{cases}
	\label{Eq:E}
\end{align}
Neglecting the electronic spin degeneracy, there is a non-degenerate ground state belonging to the $A_1$ irreducible representation of $T_d$ below a triplet of states belonging to $T_2$ and a doublet of states belonging to $E$.

In practice, we need a suitable approximation for the product of Bloch functions $\varphi^\ast_\mu(\vec{r})\varphi_\nu(\vec{r})$ and to find the right binding potential $U(\vec{r})$ for our donor.
The latter consists of the bulk-screened Coulomb potential $U_c(\vec{r})=-e^2/(4\pi\epsilon_0\epsilon_r r)$, where $e$ is the electron charge, $\epsilon_0$ is the permittivity of free space, $\epsilon_r$ the dielectric constant of the host material, and $r=|\vec{r}|$ the distance from the nucleus of the substitutional donor.

In the short-range limit, i.e. near the unit cell containing the donor atom, the approximation of a bulk-screened Coulomb potential described by $\epsilon_r$ breaks down. The impurity deforms the charge density the valence electrons it interacts with. To obtain the energies in \rfig{Fig:1}b we need to include corrections close to the impurity.

Here, we use a central cell correction (CCC) in the style proposed by Gamble \textit{et al.}\cite{gamble2015multivalley}
\begin{align}
	U_{cc}(\vec{r}) = A_0 e^{-r^2/(2a^2)}+A_1\sum_i e^{-|\vec{r}-b\vec{t}_i|^2/(2c^2)},
	\label{Eq:CC}
\end{align}
where we sum over all tetrahedral bond directions $\vec{t}_i \in \{(1,1,1),(-1,1,-1),(1,-1,-1),(-1,-1,1)\}$.
All CCC parameters need to be found via optimization (see \rsec{Sec:BasisOpti}). \rfig{Fig:1}d shows the constant energy surfaces of the CCC given in \rtab{Tab:SmallBasis}. 

The Bloch functions can be expanded in a Fourier series,
\begin{align}
	\varphi_\mu(\vec{r})=u_{\vec{k}_\mu}(\vec{r})e^{i\vec{k}_\mu\vec{r}}=e^{i\vec{k}_\mu\vec{r}}\sum_{\vec{G}}A_{\vec{G}}^\mu e^{i\vec{G}\vec{r}},
\end{align}
where the sum goes over reciprocal lattice vectors $\vec{G}$.
The Fourier coefficients $\{A^\mu_{\vec{G}}\}$ were calculated by Gamble \textit{et al.}\cite{gamble2015multivalley} using density functional theory.
Additionally, they established that the series of products of Bloch functions
\begin{align}
	{\varphi_\mu}^\ast(\vec{r})\varphi_\nu(\vec{r})=\sum_{\vec{G},\vec{G}^{\prime}}\left(A^\mu_{\vec{G}^\prime}\right)^{\ast}A^\nu_{\vec{G}}e^{i(\vec{k}_\nu-\vec{k}_\mu+\vec{G}-\vec{G}^\prime)\vec{r}}
	\label{Eq:BlochProduct}
\end{align}
is well-converged at $|\vec{G}-\vec{G}^{\prime}|\leq 4.4 \times 2\pi/\SI{0.543}{\nano\meter}$ and can be truncated at this point.

To solve \req{Eq:EMTEqs}, we expand the envelope functions $F^\mu(\vec{r})$ over a finite set of orbitals $F^\mu_a(\vec{r})$,
\begin{align}
	\psi(\vec{r}) &=\sum_{\mu}\sum_{a}C^\mu_a F^\mu_a(\vec{r})\varphi_\mu(\vec{r})\\
	&=\sum_{\mu}\sum_{a}C^\mu_a\chi^\mu_a(\vec{r}).
	\label{Eq:BasisExpansion}
\end{align}
We also define $\chi^\mu_a$ as a valley orbit basis state.
In this basis, the EMT Hamiltonian has the following form:
\begin{align}
	\begin{split}
		h^{\mu\nu}_{a,b}=&\delta_{\mu,\nu}\braket{F^\mu_a(\vec{r})|\mathbf{T}_\mu+U(\vec{r})|F^\nu_b(\vec{r})}\\
		&+\braket{F^\mu_a(\vec{r})|V^{VO}_{\mu\nu}(\vec{r})|F^\nu_b(\vec{r})}
	\end{split}
	\label{Eq:EMTHamiltonian}
\end{align}
The valley-diagonal blocks in the first line contain the kinetic energy and the impurity potential.
They are coupled via the valley-orbit elements in the second row. 
Finally, we can express \req{Eq:EMTEqs} as a generalized eigenvalue problem,
\begin{align}
	\mathbf{h}\vec{C}_i=\epsilon_i\mathbf{S}\vec{C}_i,
	\label{Eq:EMTEqsMatrixForm}
\end{align}
where $S^{\mu\nu}_{a,b}=\braket{\chi^\mu_a|\chi^\nu_b}=\delta_{\mu,\nu}\braket{F^\mu_a|F^\nu_b}$ is the block diagonal overlap matrix.

\section{Quantum Chemistry}
\label{Sec:QuantumChemistry}
Quantum chemistry is the field of science that seeks to describe the electronic structure of molecules from first principles using quantum mechanics.
This section will present the Hamiltonian of a two-donor molecule.
Then, we introduce computational chemistry methods, viz. Hartree-Fock (HF) and Configuration Interaction (CI), applied to donors in silicon.

\subsection{The donor molecule}
The Hamiltonian of the donor molecule is given by
\begin{align}
	\mathbf{H}=\mathbf{h}(1) + \mathbf{h}(2) + \mathbf{r}_{12}(1,2)
	\label{Eq:MultiElectronHamiltonian}
\end{align}
Here, $\mathbf{h}(i)$ the one-electron effective mass Hamiltonian of the $i$th electron in the field of both donors as presented in \rsec{Sec:EMT}.
It is identical for both electrons and the 1 indicates the dependence on the space and spin coordinates of electron number 1.
The electron-electron repulsion operator is given by 
\begin{align}
	\mathbf{r}_{12}(1,2)=\frac{e^2}{4\pi\epsilon_r\epsilon_0|\vec{r}_1-\Vec{r}_2|}
	\label{Eq:TwoElectronOperator}
\end{align}
where $\vec{r}_i$ is the position operator of the $i$th electron and all other quantities are the same as for the donor Coulomb potential in \rsec{Sec:EMT}.

Despite the superficial simplicity of the Hamiltonian \ref{Eq:MultiElectronHamiltonian}, the multi-electron nature of the problem renders it very complex, often resisting semi-analytic solution and typically requiring sophisticated computational methods like the ones outlined below.

\subsection{Hartree-Fock Approximation}
\label{Sec:HartreeFock}
In the Hartree-Fock (HF) model, we treat each electron as an independent particle and only consider interactions to the mean field of the other electrons.
\subsubsection{Slater Determinants}
We describe each electron by a spin orbital $\phi_i$, a one-electron wave function that is the product of a spatial orbital $\psi_i$ and the spin, which can either be up $v_{\uparrow}$ or down $v_{\downarrow}$.
In the simplest case we can have only one spin-up $\phi_{\uparrow}(1)=\psi_{\uparrow}(1)v_{\uparrow}(1)$ and one spin-down  $\phi_{\downarrow}(1)=\psi_{\downarrow}(1)v_{\downarrow}(1)$ orbital.
The many electron wave-function $\Phi(1,2,...)$ is a product of spin orbitals.
It must not only satisfy the Schr\"odinger equation, but also be antisymmetric with respect to the interchange of the space and spin coordinates of any two electrons:
\begin{align}
	\Phi(1,..,i,...,j,...,N)=-\Phi(1,..,j,...,i,...,N)
	\label{Eq:AntisymmetryPrinciple}
\end{align}
This condition can conveniently be satisfied by arranging the spin orbitals in a Slater determinant (SD). In our two-electron example this is simply given by:
\begin{align}
	\Phi(1,2)&=\frac{1}{\sqrt{2}}
	\begin{vmatrix}
		\phi_{\uparrow}(1) & \phi_{\downarrow}(1)\\
		\phi_{\uparrow}(2) & \phi_{\downarrow}(2)
	\end{vmatrix}
	\label{Eq:SDs}\\
	&=\frac{1}{\sqrt{2}}\left(\psi_{\uparrow}(1)v_{\uparrow}(1)\psi_{\downarrow}(2)v_{\downarrow}(2)-\psi_{\downarrow}(1)v_{\downarrow}(1)\psi_{\uparrow}(2)v_{\uparrow}(2)\right)\nonumber
\end{align}
Allowing each spin to have a different spatial orbital, as above, results in the Unrestricted Hartree-Fock (UHF) approximation. Restricting the spatial orbitals for each spin to be identical, i.e. $\psi_{\uparrow}=\psi_{\downarrow}$, is called Restricted Hartree-Fock (RHF).

\subsubsection{Hartree-Fock Equations}
To find the molecular orbitals (MOs) $\phi_i$ that minimize the energy of a single Slater Determinant, we need to solve a set of effective one-electron equations, the HF Equations:
\begin{align}
	\mathbf{F}_i\phi_i=\epsilon_i\phi_i.
	\label{Eq:FockEquations}
\end{align}
This is the pseudo-eigenvalue problem, as the Fock operator itself depends on the MOs as outlined below. In this case it is given by
\begin{align}
	\mathbf{F}_i=\mathbf{h}+\mathbf{J}_{\uparrow}+\mathbf{J}_{\downarrow}-\mathbf{K}_i.
	\label{Eq:FockOperator}
\end{align}
Here, we introduced the Coulomb $\mathbf{J}_i$ and exchange $\mathbf{K}_i$ operators, which we define via the following integrals
\begin{align}
	\braket{\phi_j|\mathbf{J}_i|\phi_j}&=\braket{\phi_i(1)\phi_j(2)|\mathbf{r}_{12}|\phi_i(1)\phi_j(2)}\\
	\braket{\phi_j|\mathbf{K}_i|\phi_j}&=\braket{\phi_i(1)\phi_j(2)|\mathbf{r}_{12}|\phi_j(1)\phi_i(2)}.
\end{align}
The integral over the Coulomb operator is just the classical repulsion between the charge distributions $\phi_j^2$ and $\phi_i^2$.
Note how the $\mathbf{K}$ operator exchanges the two orbitals on the right-hand side.
The exchange integral has no classical analogy.
Intuitively, the Fock operator can be interpreted as follows:
The up spin electron sees the Coulomb potential from electrons in spin up and down orbitals plus the exchange potential from the spin up orbital.
There is no exchange interaction between electrons of different spin.
In RHF, the Fock operator is equal for both spin orbitals and \req{Eq:FockOperator} simplifies to
\begin{align}
	\mathbf{F}=\mathbf{h}+2\mathbf{J}-\mathbf{K}.
\end{align}
We will focus on this case in the following.
\subsubsection{Self-Consistent Field Iterations}
The Fock operator itself depends on the orbitals $\phi_i$ and \req{Eq:FockEquations} can only be solved iteratively.

In practice, the MOs are expanded over a finite orbital basis set, just like the EMT envelope functions in \req{Eq:BasisExpansion}. The matrix representation of \req{Eq:FockEquations} in the valley-orbit basis yields the Roothaan-Hall equations
\begin{align}
	\mathbf{F}\vec{C}_i=\epsilon_i\mathbf{S}\vec{C}_i,
	\label{Eq:RoothaanHall}
\end{align}
the generalized eigenvalue problem, where $S^{\mu,\nu}_{a,b}=\braket{\chi^\mu_a|\chi^\nu_b}$ and $F^{\mu,\nu}_{a,b}=\braket{\chi^\mu_a|\mathbf{F}|\chi^\nu_b}$ are $\mu$-$\nu$-valley matrix blocks of the Overlap and Fock matrix respectively.
The first step to self-consistently solve \req{Eq:RoothaanHall} is to precompute the two-electron integrals.
They are approximately valley local\cite{koiller2001exchange,wellard2003electron,saraiva2011intervalley},
\begin{align}
	\begin{split}
		\braket{\chi^\mu_a\chi^\kappa_c|\mathbf{r}_{12}|\chi^\nu_b\chi^\lambda_d}&=\delta_{\mu\nu}\delta_{\kappa\lambda}\braket{F^\mu_a F^\kappa_c|\mathbf{r}_{12}|F^\mu_b F^\kappa_d}\\
		&=\delta_{\mu\nu}\delta_{\kappa\lambda}(ab|cd)_{\mu\kappa},
	\end{split}
\end{align}
as the fast oscillatory part of the Bloch function,
\begin{align}
	\varphi^\ast_\mu(1)\varphi_\nu(1)=u_{-\vec{k}_\mu}(\vec{r}_1)u_{\vec{k}_\nu}(\vec{r}_1)e^{i(\vec{k}_\nu-\vec{k}_\mu)\vec{r}_1},
\end{align}
suppresses the contribution of mixed valley charge distributions, i.e. for $\vec{k}_\mu\neq\vec{k}_\nu$.
We also introduced the four-index tensor $(ab|cd)_{\mu\kappa}$, a convenient chemistry notation.
Given an initial orbital $\psi_i(1)=\sum_\mu\sum_a C^\mu_{a,i}\chi^\mu_a(1)$, we find
\begin{align}
	&\braket{\chi^\mu_a|\mathbf{J}|\chi^\nu_b}
	=\delta_{\mu\nu}\sum_{\kappa}\sum_{c,d}{C^\kappa_{d,i}}^\ast C^\kappa_{d,i}(ab|cd)_{\mu\nu},
	\label{Eq:Jmat}\\
	&\braket{\chi^\mu_a|\mathbf{K}|\chi^\nu_b}
	=\sum_{c,d}{C^\mu_{c,i}}^\ast C^\nu_{d,i}(ad|cb)_{\mu\nu}.
	\label{Eq:Kmat}
\end{align}
Note that the J matrix is valley diagonal. It can be interpreted as the mutual electrostatic repulsion between the classical charge densities in each valley.

The self-consistent field iterations now go as follows: (i) Obtain an inital guess for the state $\psi_0$. (ii) Form the Fock matrix by application of \req{Eq:Jmat} and \req{Eq:Kmat}. (iii) Diagonalize the Fock matrix, i.e. solve \req{Eq:RoothaanHall}. (iv) Find the new ground state coefficients $C^\mu_{a,0}$ and iterate from step (ii) until the cycle converges.

\begin{figure}[htb!]
	\includegraphics[width=1.0\columnwidth]{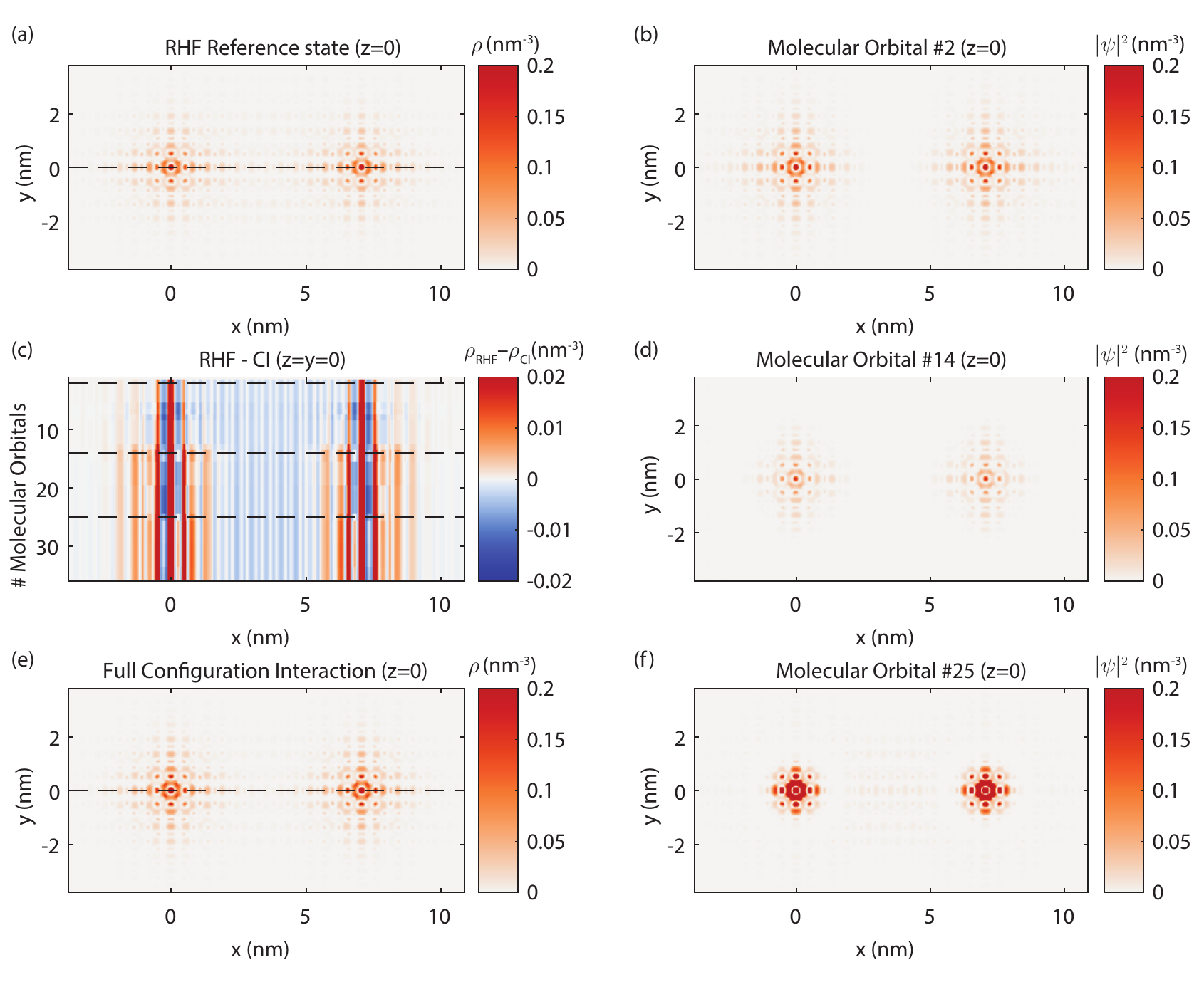}
	\caption{Illustration of the Configuration Interaction (CI) method for the exemplary case of the singlet ground state of two phosphorus donors spaced by $d=\SI{7.059}{\nano\meter}$ along [100]. The upper panel ((b), (d) and (f) offline) shows the converged molecular orbitals (MO) $|\psi_i(x,y,z=0)|^2$ for the basis set in \rtab{Tab:SmallBasis}, where the slider iterates through $i$. In the lower panel, the we perform the CI calculation including configurations in \req{Eq:SingletSAC} up to and including the MO in the top panel. We show the resulting electron density $\rho(\vec{r}_1)=\int|\Psi^{S}(\vec{r}_1,\vec{r}_2)|^2 d^3\vec{r}_2$. Only including the first MO ((a) offline) corresponds to the Restricted Hartree Fock (RHF) reference state, while including all MOs is the Full CI result ((e) offline). Adding more MOs allows the CI to account better for the electron correlation and lowers the CI energy. The radio button optionally shows the shift of the CI electron density compared to the RHF reference state. (Offline: Panel (c) gives an $z=y=0$ linecut as a function of included number of MOs on the y-axis)  }
	\label{Fig:2}
\end{figure}

The upper panel of \rfig{Fig:2} shows the fully converged $|\psi_i(x,y,z=0)|^2$ for two phosphorus donors spaced by $d=\SI{7.059}{\nano\meter}$ along [100]. The basis used is defined in \rtab{Tab:SmallBasis} and contains 3 orbitals per valley per donor resulting in a 36-dimensional problem.

\subsection{Configuration Interaction}
The HF method finds the energetically best solution a single SD can offer. However, this underestimates how the motion of the electrons influence each other, i.e. the electron correlation (EC). The Configuration Interaction (CI) method accounts for the EC energy by expanding the full multi-electron wave function $\Psi$ over a set of SDs
\begin{align}
	\Psi = \sum_i a_i \Phi_i.
\end{align}
Each of these SDs represents an electron configuration. The full multi-electron Hamiltonian (\req{Eq:MultiElectronHamiltonian}) is expressed in this basis.
By diagonalization, we determine how these configurations interact.
Including all possible configurations is called Full CI and it is the optimal solution in a given basis set. 

\subsubsection{Spin-Adapted Configurations}
Our Hamiltonian in \req{Eq:MultiElectronHamiltonian} does not contain any spin operators and $\mathbf{S}^2$ and $\mathbf{S}_z$ commute with $\mathbf{H}$.
As a result, the exact eigenfunctions of $\mathbf{H}$ are also eigenfunctions of the spin operators.
In the two-electron case, these are the well-known singlet $S$ ($\mathbf{S}^2=0$) and triplet $T$ ($\mathbf{S}^2=2$) spin functions:
\begin{align}
	S(1,2) &= \frac{1}{\sqrt{2}}(v_{\uparrow}(1)v_{\downarrow}(2)-v_{\downarrow}(1)v_{\uparrow}(2))
	\label{Eq:SingletSpin}\\
	T(1,2) &= 
	\begin{cases}
		v_{\uparrow}(1)v_{\uparrow}(2)\\
		\frac{1}{\sqrt{2}}(v_{\uparrow}(1)v_{\downarrow}(2)+v_{\downarrow}(1)v_{\uparrow}(2))\\
		v_{\downarrow}(1)v_{\downarrow}(2)
	\end{cases}
	\label{Eq:TripletSpin}
\end{align}
In the previous section, we solved the RHF Roothaan's equations and found a set of MOs $\psi_i$. 
The RHF ground state configuration was simply given by $\psi_\uparrow=\psi_\downarrow=\psi_0$ in \req{Eq:SDs}, which corresponds to a spin singlet as expected.
In principle, we can build a set of excited SDs $\Phi_{ij}$ by promoting the up and down spin orbitals to excited MOs, $\psi_i$ and $\psi_j$.
However, configurations $i\neq j$ are not eigenfunctions of the total spin operator $\mathbf{S}^2$.
We use spin-adapted configurations (SACs) instead, which conveniently split $\mathbf{H}$ into singlet and triplet blocks.
The spatial part of the singlet SACs are
\begin{align}
	\Phi^S_{ij}(1,2)=
	\begin{cases}
		\psi_i(1)\psi_j(2) & i=j\\
		\frac{1}{\sqrt{2}}(\psi_i(1)\psi_j(2)+\psi_j(1)\psi_i(2))  & i\neq j
	\end{cases}
	\label{Eq:SingletSAC}
\end{align}
and for the triplet
\begin{align}
	\Phi^T_{ij}(1,2)=\frac{1}{\sqrt{2}}(\psi_i(1)\psi_j(2)-\psi_j(1)\psi_i(2)),
\end{align}
where the latter requires $i\neq j$. Together with the spin functions in \req{Eq:SingletSpin} and \req{Eq:TripletSpin} they form the orthonormal basis in which we express our two-electron Hamiltonian:
\begin{align}
	\Psi^S(1,2)=\sum_{i\leq j} a^S_{ij} \Phi^S_{ij}(1,2)\, S(1,2)
	\label{Eq:SingletCI}\\
	\Psi^T(1,2)=\sum_{i< j} a^T_{ij} \Phi^T_{ij}(1,2)\, T(1,2)\
	\label{Eq:TripletCI}
\end{align}
Note that using Full CI, it does not matter if we employ solutions of \req{Eq:EMTEqsMatrixForm} or \req{Eq:RoothaanHall}.

\subsubsection{CI matrix elements}
All we have to do do now is to express \req{Eq:MultiElectronHamiltonian} in the singlet and triplet SAC basis.
The first step is to express the one- and two-electron integrals in the MO basis
\begin{align}
	h_{ij}&=\braket{\psi_i|\mathbf{h}|\psi_j}=\sum_{\mu,\nu}\sum_{a,b}{C^\mu_{a,i}}^\ast C^\nu_{b,j}h^{\mu\nu}_{ab}\\
	(ij|kl)&=\sum_{\mu,\nu}\sum_{a,b,c,d}{C^\mu_{a,i}}^\ast C^\mu_{b,j}{C^\nu_{c,k}}^\ast C^\nu_{d,l}(ab|cd)_{\mu\nu}.
	\label{Eq:MOtransfrom}
\end{align}
Performing the transformation of the four-index tensor one index at a time reduces the complexity to $\mathcal{O}(N_{\rm basis}^5)$.
With this, we can easily compute the the SAC matrix elements, which in the triplet case they are given by
\begin{align}
	\begin{split}
		\mathbf{H}^T=\braket{\Phi^T_{ij}|\mathbf{H}|\Phi^T_{kl}}  =&\delta_{jl}h_{ik}-\delta_{jk}h_{il}-\delta_{il}h_{jk}+\delta_{ik}h_{jl}\\
		&+(ik|jl)-(il|jk)
	\end{split}
\end{align}
The singlet elements $\mathbf{H}^S$ are a little more complex due to the cases $i=j$ in \req{Eq:SingletSAC}. By solving the eigenvalue problem
\begin{align}
	\mathbf{H}^{S/T} \vec{a}_{n}^{S/T} = E_{n}^{S/T} \vec{a}_{n}^{S/T},
\end{align}
we find the singlet/triplet $E_{n}^{S/T}$ energies as well as the corresponding coefficients $a^{S/T}_{ij}$ (\req{Eq:SingletCI} /\req{Eq:TripletCI}) of the $n$th singlet/triplet state.

In the lower panel of \rfig{Fig:2} we perform the CI calculations including configurations in \req{Eq:SingletSAC} up to and including the MO in the top panel. The resulting electron density 
\begin{align}
	\rho(\vec{r}_1)=\int |\Psi^S(\vec{r}_1,\vec{r}_2)|^2 d^3\vec{r}_2
	\label{Eq:ElectronDensity}
\end{align}
is shown for $z=0$. Including only a single MO is the RHF solution. Having only one orbital to work with, the RHF solution spreads it over both donors and introduces unnaturally high contributions from the $E$ valley-orbit states, which are wide along the axis joining the donor. The resulting energy of $E_{\rm RHF}=\SI{-98.50}{\milli\electronvolt}$ is consequently quite high. Adding only the first excited MO shifts the electron density onto the donors, reducing the $E$-states contributions and lowering the energy to $E_{\rm CI}=\SI{-107.03}{\milli\electronvolt}$. Adding more MOs increases this effect. The electron density of the Full CI solution, i.e. including all 36 MOs, looks much more like the expected solution of two $A_1$-like states\cite{gamble2015multivalley}. The resulting energy is lower at $E_{\rm CI}=\SI{-109.64}{\milli\electronvolt}$. The energy difference between RHF and Full CI is called correlation energy and notably accounts for $\SI{10}{\percent}$ of the total energy within this basis set.

\section{Basis Sets}
\label{Sec:BasisOpti}

To solve the EMT equations and to find our MOs, we expanded our envelope functions over a set of valley-orbit basis functions $F^\mu_a(\vec{r})$ (\req{Eq:BasisExpansion}).
In practice, it is impossible to use a complete basis, as we can only work with a finite number of functions.
The size and the type of the basis we chose will determine the level of accuracy we can achieve.

Ideally, every single basis function already reproduces the shape of the wave function we would like to model.
In our case, these are Slater type orbitals (STOs), which mimic the exact orbitals for the hydrogen atom.
In the silicon crystal, the STOs are anisotropic\cite{voisin2020valley}; a $+z$-valley orbital centered at the origin has the form:
\begin{align}
	\text{STO}(\vec{r};\mathcal{N},n_x, n_y, n_z,\alpha_{\perp},\alpha_{\parallel})=\mathcal{N} x^{n_x} y^{n_y} z^{n_z} e^{-\sqrt{\alpha_{\perp}(x^2+y^2)+\alpha_{\parallel}z^2}},
\end{align}
where $\mathcal{N}$ is a normalization factor, the $n_i$ are the exponents of polynomial prefactors and the $\alpha$ the anisotropic decay constants.

Gaussian type orbitals (GTOs) allow for efficient evaluation of the one and two-electron integrals. The the one- and two-centered Coulomb kernels can be accurately decomposed into a compact Gaussian quadrature. In a GTO basis, the resulting matrix elements are then a sum of separable integrals over Gaussian functions, which can be computed analytically\cite{shankar2021computational}.
The GTO analogue to the STO above has the following shape
\begin{align}
	\text{GTO}(\vec{r};\mathcal{N},n_x, n_y, n_z,\alpha_{\perp},\alpha_{\parallel})=\mathcal{N} x^{n_x} y^{n_y} z^{n_z} e^{-\alpha_{\perp}(x^2+y^2)-\alpha_{\parallel}z^2}.
\end{align}
The GTOs have a zero derivative at the center and are missing the cusp of a STO, making it hard to predict the proper behaviour close to the nucleus.
Additionally, GTOs have a steeper flank.
As a rule of thumb one needs three times as many GTOs as STOs to achieve the same accuracy\cite{jensen2017introduction}. This makes the transformation in \req{Eq:MOtransfrom}, which scales as $\mathcal{O}(N_{\rm basis}^5)$, particularly costly in a GTO basis. The standard solution to this problem is to fit a number of $n$ GTOs to a STO and contract them to represent one orbital.
A single STO-$n$G basis function then has the following form
\begin{align}
	\text{STO-$n$G}(\vec{r};\mathcal{N},n_x, n_y, n_z,\alpha_{\perp},\alpha_{\parallel})=\mathcal{N}\sum_{i=1}^{n} \text{GTO}(\vec{r};\mathcal{N}_i,n_x, n_y, n_z,\alpha_{\perp}\cdot\beta_i,\alpha_{\parallel}\cdot\beta_i),
	\label{Eq:STOnG}
\end{align}
where the $\mathcal{N}_i$ and $\beta_i$ are the fit parameters (see \rfig{Fig:App1}).
We choose $n=3$ as adding more GTOs to the fit gives little improvement. The basis functions in the 5 other valleys are found via permutation of the exponents and polynomials.

\begin{table}[htb!]
	\caption{Small basis set parameters resulting in the energies on the right. The last orbital is added to fit the $D^-$ energy.}
	\label{Tab:SmallBasis}
	\begin{minipage}[t]{.49\linewidth}
		Basis Parameters\\
		\begin{tabular}{llll}
			\hline
			\hline
			\#& $(n_x,n_y,n_z)$& $\alpha_{\perp} ({\rm nm}^{-2})$ & $\alpha_{\parallel}({\rm nm}^{-2})$   \\
			\hline
			1& $(0,0,0)$ & 0.246 & 0.950  \\
			2& $(0,0,0)$ & 2.408 & 5.458  \\
			\hline
			3& $(0,0,0)$ & 0.0944 & 0.478
		\end{tabular}
		\ \\
		\ \\
		Central Cell Parameters\\
		\begin{tabular}{p{0.15\linewidth}p{0.45\linewidth}}
			\hline
			\hline
			$A_0$& $\SI{-1.395}{\milli\electronvolt}$\\
			$A_1$& $\SI{-2717.0}{\milli\electronvolt}$\\
			$a$& $\phantom{-}\SI{0.127}{\nano\meter}$\\
			$b$& $\phantom{-}\SI{0.194}{\nano\meter}$\\
			$c$& $\phantom{-}\SI{0.0972}{\nano\meter}$
		\end{tabular}
	\end{minipage}
	\begin{minipage}[t]{.49\linewidth}
		\centering
		Energy Levels\\
		\includegraphics[]{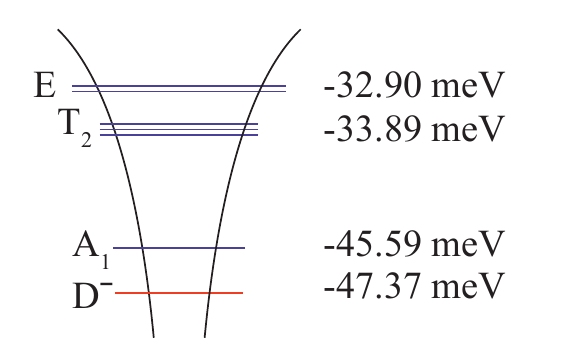}
	\end{minipage}
\end{table}

Having decided on the style of basis, we need to find a set of STO-$n$Gs that models all features of the two-donor system with sufficient accuracy.
We follow Gamble \textit{et al.}\cite{gamble2015multivalley} and variationally fix a CCC and basis that reproduces the energy levels of a single phosphorus donor.
We try a small basis (SB) containing only $1s$-orbitals ($n_x=n_y
=n_z=0$). The CCC in \rtab{Tab:SmallBasis} together with the orbitals \#1 to \#2 produce the D$^0$ energies on the right.
The SB shows good agreement in the $A_1$ ground state. However, it was not possible to match the E and T$_2$ energies at the same time.

We proceed to test a larger basis (LB) that additionally contains  $2s$-orbitals ($n_x=n_y
=n_z=2$).
The CCC and orbitals \#1 to \#5 in \rtab{Tab:LargeBasis} achieve good accuracy for all D$^0$ valley configurations.

To be able to model configurations with two electrons loaded on one donor, we optimize an additional orbital to produce the D$^-$ energy.
We find that it is sufficient to add orbital \#3 to the SB and \#6 to the LB to achieve the D$^-$ energies on the right.

\begin{table}[htb!]
	\caption{Large basis set parameters resulting in the energies on the right. The last orbital is added to fit the $D^-$ energy.}
	\label{Tab:LargeBasis}
	\begin{minipage}[t]{.49\linewidth}
		Basis Parameters\\
		\begin{tabular}{llll}
			\hline
			\hline
			\#& $(n_x,n_y,n_z)$& $\alpha_{\perp} ({\rm nm}^{-2})$ & $\alpha_{\parallel}({\rm nm}^{-2})$   \\
			\hline
			1& $(0,0,0)$ & 0.227 & 0.864  \\
			2& $(0,0,0)$ & 1.121 & 3.406  \\
			3& $(2,0,0)$ & 17.04 & 103.9  \\
			4& $(0,2,0)$ & 17.04 & 103.9  \\
			5& $(0,0,2)$ & 17.04 & 103.9  \\
			\hline
			6& $(0,0,0)$ & 0.0695 & 0.302  \\
		\end{tabular}
		\ \\
		\ \\
		Central Cell Parameters\\
		\begin{tabular}{p{0.15\linewidth}p{0.45\linewidth}}
			\hline
			\hline
			$A_0$& $\SI{-1.626}{\milli\electronvolt}$\\
			$A_1$& $\SI{-2330.8}{\milli\electronvolt}$\\
			$a$& $\phantom{-}\SI{0.140}{\nano\meter}$\\
			$b$& $\phantom{-}\SI{0.210}{\nano\meter}$\\
			$c$& $\phantom{-}\SI{0.102}{\nano\meter}$
		\end{tabular}
	\end{minipage}
	\begin{minipage}[t]{.49\linewidth}
		\centering
		Energy Levels\\
		\includegraphics[]{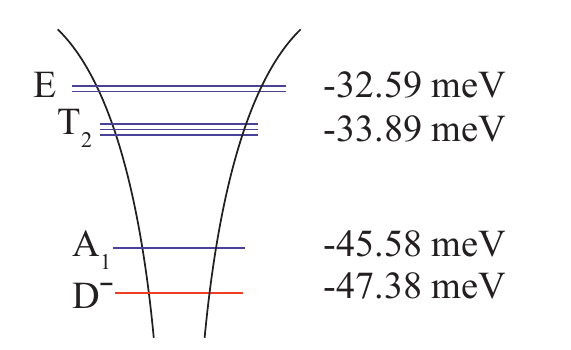}
	\end{minipage}
\end{table}

\begin{figure}[htb!]
	\includegraphics[width=1.0\columnwidth]{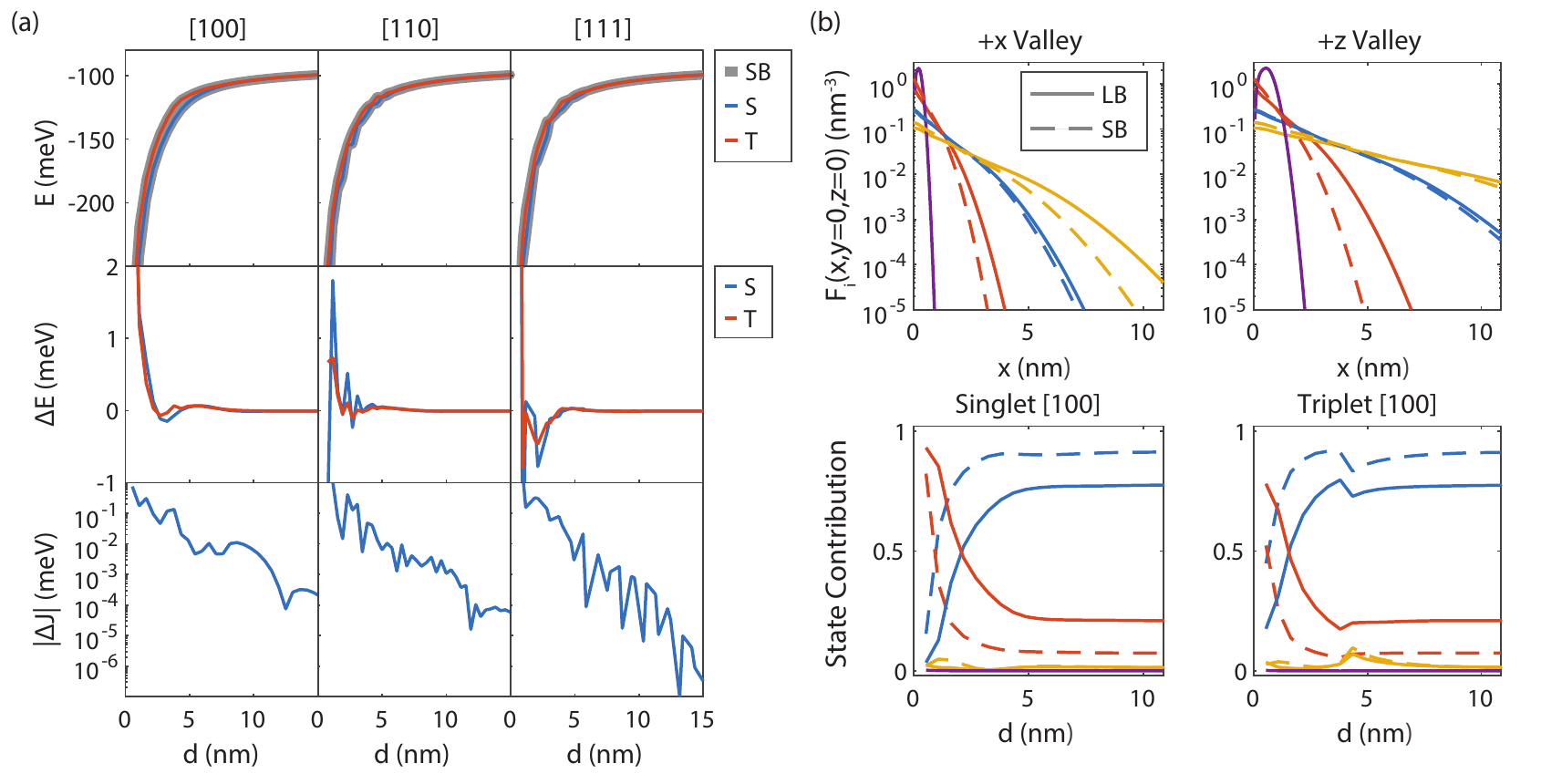}
	\caption{Basis Set analysis. (a) Direct comparison of the Large Basis (LB in \rtab{Tab:LargeBasis}) and Small Basis (SB in \rtab{Tab:SmallBasis}) results as a function of donor separation along three crystal directions. The first row shows the lowest singlet (S) and triplet (T) energies for the LB and SB results as a grey shadow. The second shows the energy difference $\Delta E$ between the LB and SB for the S and T respectively. Both converge to the same value, i.e . $\Delta E=0$ for large $d$.  The third is the resulting difference in exchange interaction strength on a log scale. (b) The first row shows a $y=z=0$  line cut of the LB (full) and SB (dashed) anisotropic basis functions along (left) and perpendicular (right) to the valley orientation.  For the SB the blue, red and yellow curves belong to the orbitals \#1, \#2 and \#3  in \rtab{Tab:SmallBasis}, while for the LB they belong to the orbitals \#1, \#2 and \#6  in \rtab{Tab:LargeBasis}. Note that for the LB the $n_z=n_y=2$ states vanish in the chosen line cut and only orbital \#3 is shown in purple. The contributions of these basis states (\req{Eq:StateContribution}) to the lowest singlet and triplet states as a function of donor separation along [100] are shown in the bottom panels. Here we use the same colour code as in the top panels, e.g. the contribution of the state function visualized by a full blue line in the top panels, belong to the full blue line curves in the bottom panels. However, the purple curve accounts for all $2s$-orbitals.}
	\label{Fig:3}
\end{figure}

In the following, we will test our two basis sets on two exchange-coupled donors and compare the results to literature to get a better understanding of their limitations.
The first row of \rfig{Fig:3}a shows the LB results for lowest singlet and triplet energy states as a function of donor separation along the [100], [110] and [111] crystal directions, in comparison to the SB results as a gray shade. The differences between both basis sets are very subtle. For both sets, the energies approach twice the energy of an isolated donor for large distances, while the singlet-triplet splitting decays. The second row shows the difference $\Delta E$ between the singlet/triplet energies  calculated using the SB and the LB, while the third row compares the exchange interaction strength
\begin{align}
	J = E^{T}_0- E^{S}_0,
	\label{Eq:Exchange}
\end{align}
on a logarithmic scale. The sensitivity to the choice of basis set decreases with distance and the singlet and triplet show a similar trend in sensitivity, i.e. amplitude of $\Delta E$. As a function of distance, the singlet and triplet alternate in being more sensitive, but converge to the same value, i.e . $\Delta E=0$ for large $d$. The shape of the basis functions and especially the overlap of the flanks determine these fine features.

The first row of \rfig{Fig:3}b shows a $y=z=0$ line cut of the $+x$- and $+z$-valley LB (full lines) and SB (dashed lines) basis functions in the top panel, i.e. the line cuts parallel and perpendicular to the valley direction.
The bottom panel gives their contributions to the lowest singlet and triplet states for two donors separated by $d$ along the [100] crystal axis in the same colorcode.
The state contributions (SC) of the $a$th basis state are computed as
\begin{align}
	SC_a(\Psi^{S/T}) = \sum_\mu\int|\braket{F^\mu_a(\vec{r}_1)\varphi_\mu(\vec{r}_1)|\Psi^{S/T}(\vec{r}_1,\vec{r}_2)}|^2 d^3\vec{r}_2,
	\label{Eq:StateContribution}
\end{align}
where $F^\mu_a$ is the $a$th basis state in the $\mu$th valley, which differs from the $a$th basis state in another valley through the permutation of the decay constants and polynomial prefactors.
The SB $1s$-orbitals have faster decaying flanks, i.e. the LB orbitals are flatter.
As a result, the LB states put more support on the faster decaying red orbital to offset the steeper flank of the primary (blue lines in \rfig{Fig:3}) orbital.
Note that although the $2s$-orbital contributions are small, they have high amplitudes at the CCC and are crucial to get the right valley energies.
For $d\lessapprox\SI{5}{\nano\meter}$ the donor molecule is not dissociated yet. 
The narrow (red lines in \rfig{Fig:3}) orbitals dominate the wave function. This can be understood considering the hydrogen atom. The ground state electronic wavefunction has a radius of $r=a_{\rm B}=\SI{0.53}{\angstrom}$. When we bring two hydrogen atoms together to zero distance, we get the limiting case of a helium atom, which has a narrower radius of $r=\SI{0.31}{\angstrom}$. Equally the short distance limit of a donor molecule is narrower, leading to the greater contributions of red orbitals. Both electrons are spread over the increasingly helium-like donor cluster. Neither our CCC nor our basis is designed for this limit, resulting in larger $\Delta E$.
For larger distances, the system acts more like two separate donors with one $A_1$ electron on each (see \rsec{Sec:ExchangeCoupledDonors}).
Their energies have been successfully optimized for both basis sets, resulting in $\Delta E$ nicely converging to zero.

\begin{figure}[htb!]
	\centering
	\includegraphics[width=0.6\columnwidth]{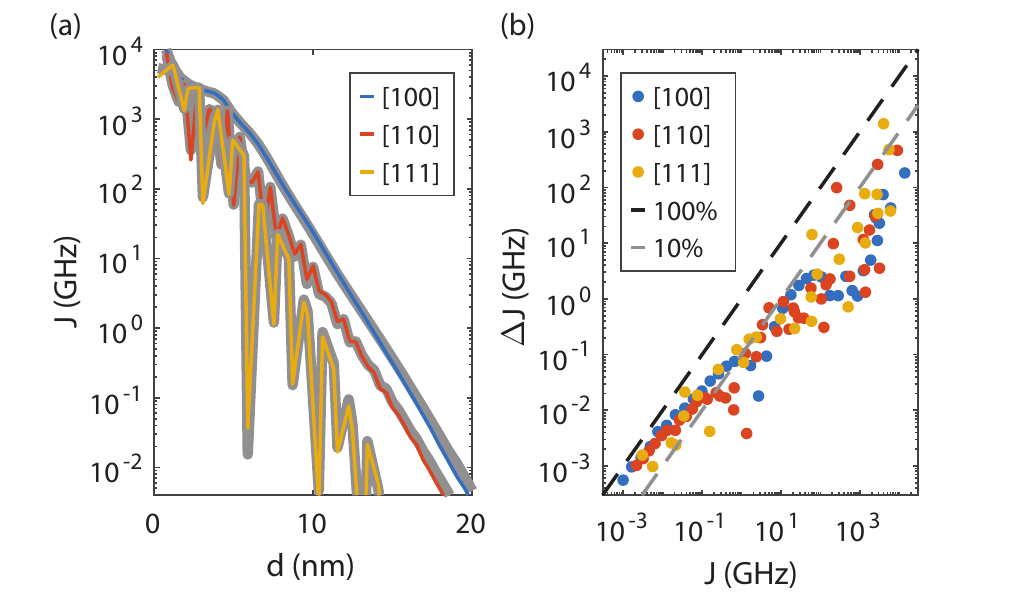}
	\caption{(a) Comparison between the small basis (SB) and large basis (LB) Full CI exchange interaction strength results in frequency units as a function of donor separation along the [100], [110] and [111] crystal directions. The SB results are shown as a grey shadow. (b) Double logarithmic plot of the difference in exchange interaction strength between SB and LB sets as a function of the exchange strength. The diagonal dashed lines mark the $\Delta J/J=100\%$ and $10\%$ regions.}
	\label{Fig:4}
\end{figure}
\rfig{Fig:4}a gives our Full CI multi-valley EMT results for the LB (colored) and SB (grey shade).
The curves show exponential decay for $d>\SI{5}{\nano\meter}$ and the typical valley oscillations for crystal orientations not in $\braket{100}$.
Additionally, $J_{[100]}(d)>J_{[110]}(d)>J_{[111]}(d)$ due to the effective mass anisotropy (see \rsec{Sec:ExchangeCoupledDonors}).
The LB exchange interaction strength decays slightly faster due to the flatter nature of the relevant basis states (red and blue) that makes the difference between the bonding and antibonding states less pronounced.
We conclude that for the study of two exchange-coupled donors with separation $d>\SI{5}{\nano\meter}$ both basis sets are equally viable.
\rfig{Fig:4}b shows the difference $\Delta J$ as a function of $J$. The diagonal dashed lines mark the $\Delta J/J=100\%$ and $10\%$ regimes. We observe that it is very difficult to precisely compute very small values of $J$. This is a natural effect and will happen in all types of calculations. 

Finally, let us consider the computational cost. The simulations were performed on the Gadi supercomputer\cite{Gadi}. A LB iteration typically costs $\SI{120}{SU}$ (`Service Units' or computer core hours), while a SB step costs $\SI{4.7}{SU}$. We here showed that the SB is sufficient for $d\gtrapprox\SI{5}{\nano\meter}$ and we can use it to perform exhaustive 3D iterations (see \rfig{Fig:6}) including $N\approx 8500$ iterations run with approximately $\SI{40}{kSU}$. The carbon footprint of all simulations run in this project is calculated in \rapp{App:Carbon}.

\section{Study of exchange-coupled donors}
\label{Sec:ExchangeCoupledDonors}
\begin{figure}[htb!]
	\includegraphics[width=1.0\columnwidth]{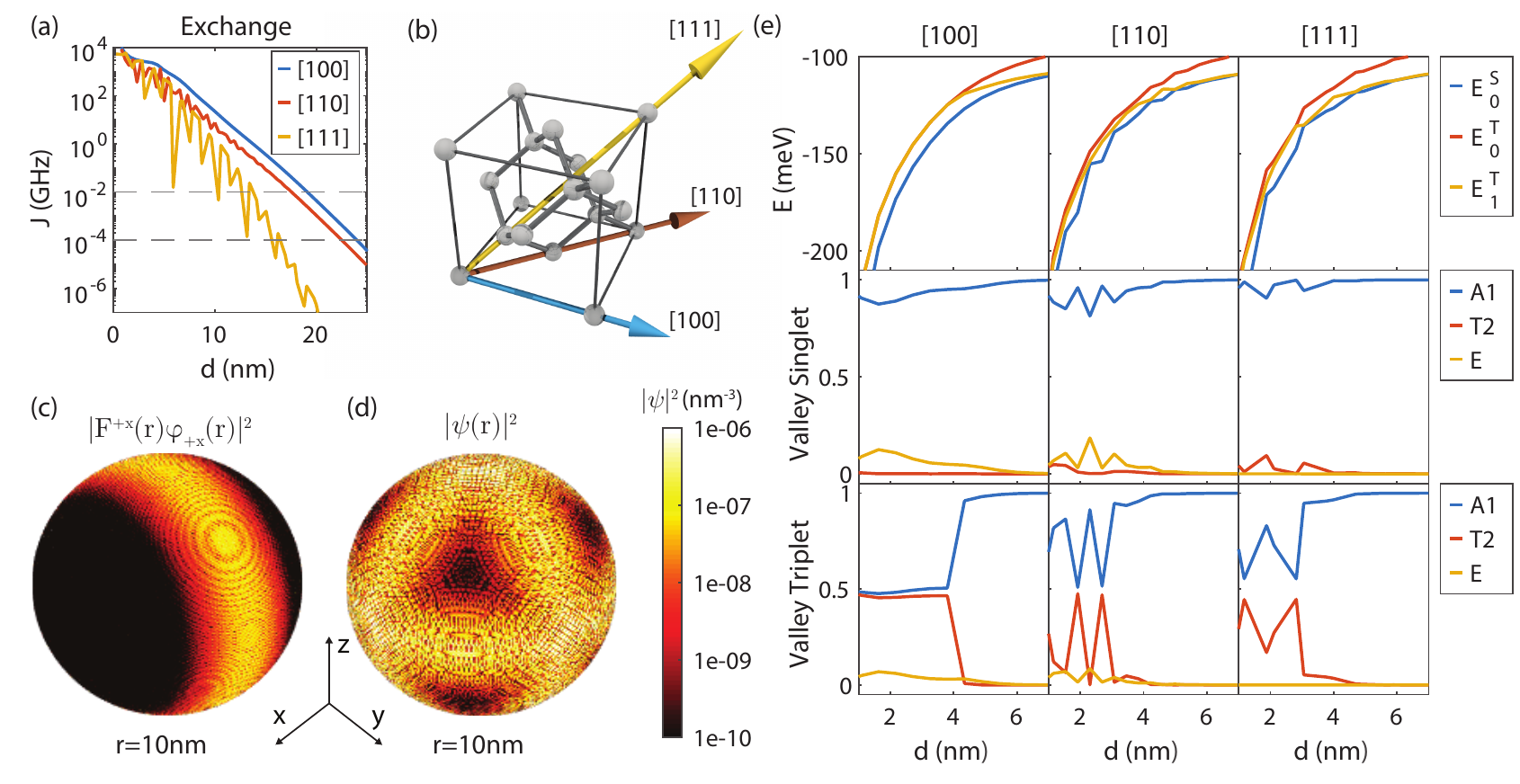}
	\caption{(a) Exchange interaction strength $J = E^{T}_0- E^{S}_0$ as a function of donor separation along three crystal orientations marked by the correspondingly colored arrow in the silicon unit cell in (b). The dashed lines indicate the exchange interaction strength range required for high-fidelity CROT gates. (c) The $+x$-valley contribution $|F^{+x}(\vec{r})\varphi_{+x}(\vec{r})|^2$ to the full $A_1$ ground state $|\psi(\vec{r})|^2$ in (d). They are plotted on a log scale on the surface of a sphere with $r=\SI{10}{\nano\meter}$ around the donor. (e) Analysis panel as function of distance for the crystal orientations in (b). The first row gives the lowest three energy states, a singlet and two triplets. The second/third row give the valley contributions computed via \req{Eq:ValleyContribution} for the lowest singlet and triplet. At roughly $d\approx \SI{4}{\nano\meter}$, the triplet has an avoided crossing and gains $T_2$ contributions.}
	\label{Fig:5}
\end{figure}

We now have an appropriate CCC and basis set and are ready use our Full CI code to study two exchange-coupled donors. \rfig{Fig:5}a shows the exchange interaction strength as a function of donor separation along different crystal axes sketched in \rfig{Fig:5}b.
\begin{figure}[htb!]
	\includegraphics[width=1.0\columnwidth]{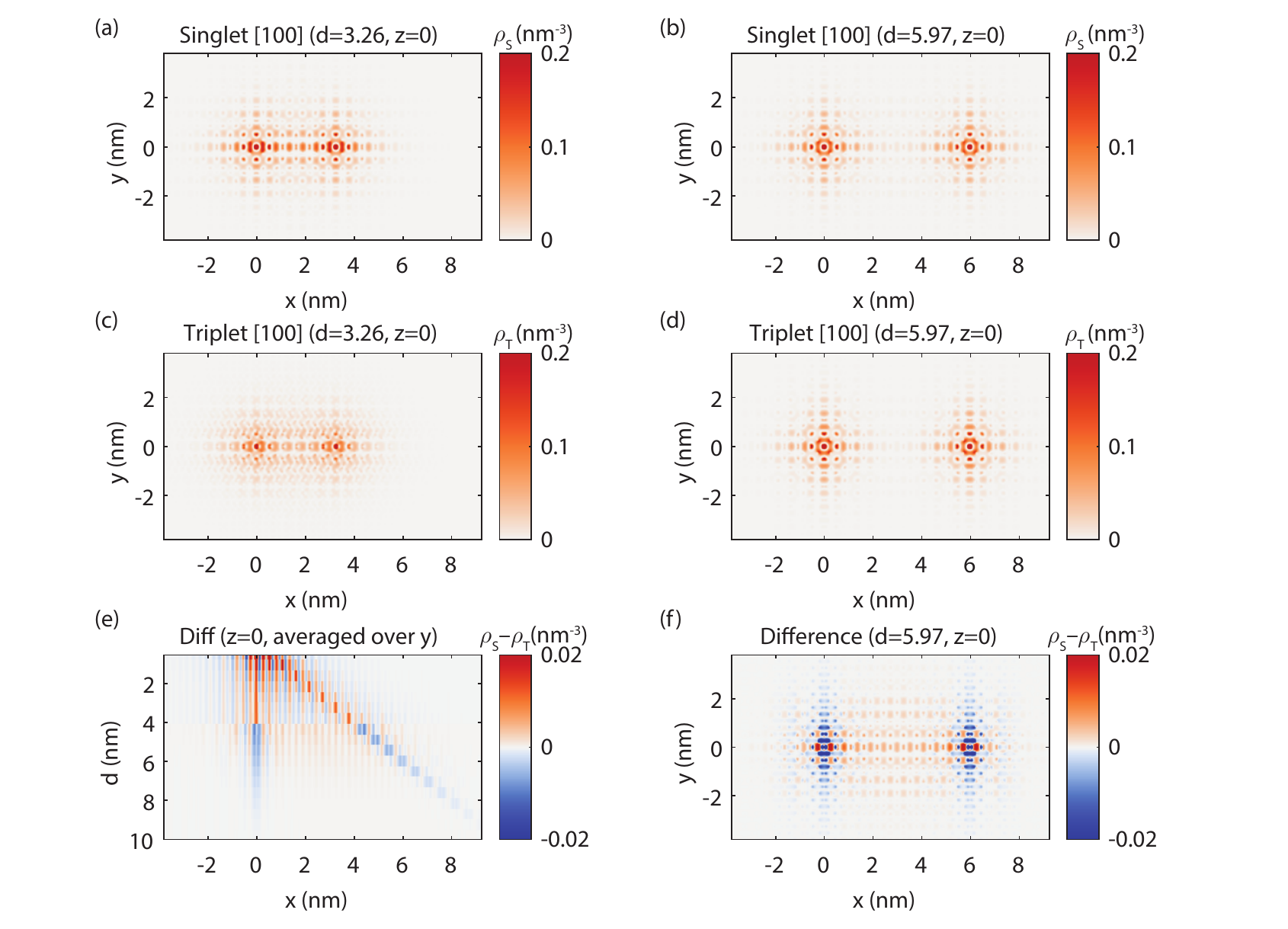}
	\caption{Electron density \req{Eq:ElectronDensity} calculations for one donor placed at the origin and one moved along [100] (x-axis) with the slider. Shown is a $z=0$ cutplane. The radio-button selects between the electron density of the lowest singlet ((a) and (b) offline), triplet  ((c) and (d) offline) or their difference  ((e) and (f) offline. (e) averages over the $y$-axis and iterates over all donor distances $d$ instead).}
	\label{Fig:6}
\end{figure}
\begin{figure}[htb!]
	\includegraphics[width=1.0\columnwidth]{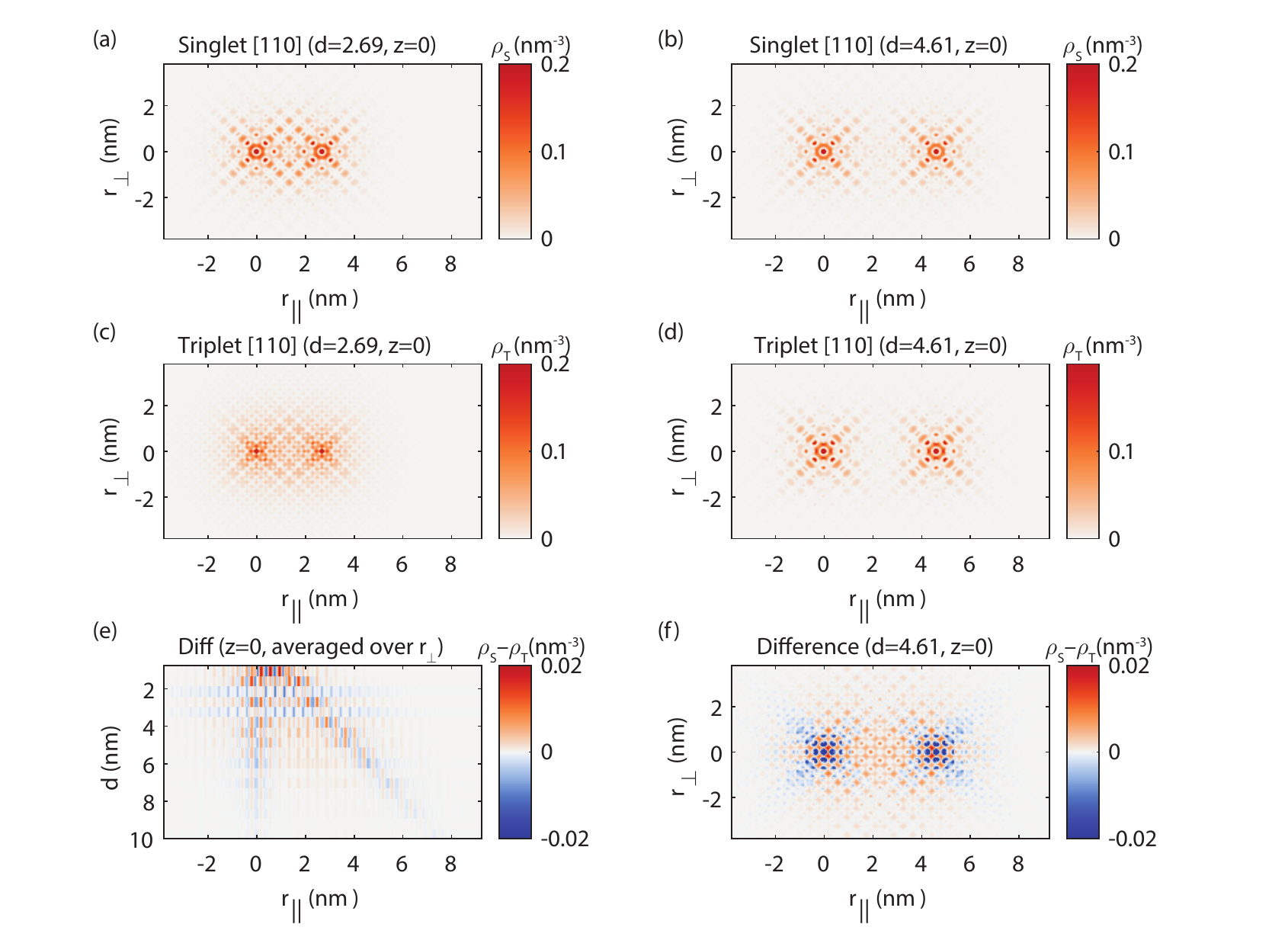}
	\caption{Electron density \req{Eq:ElectronDensity} calculations for one donor placed at the origin and the second moved along [110] with the slider. Shown is a $z=0$ cutplane with $r_{\parallel}$ being the $x$-$y$-bisector. The radio-button selects between the electron density of the lowest singlet ((a) and (b) offline), triplet  ((c) and (d) offline) or their difference  ((e) and (f) offline. (e) averages over the $r_{\perp}$-axis and iterates over all donor distances $d$ instead).}
	\label{Fig:7}
\end{figure}
\begin{figure}[htb!]
	\includegraphics[width=1.0\columnwidth]{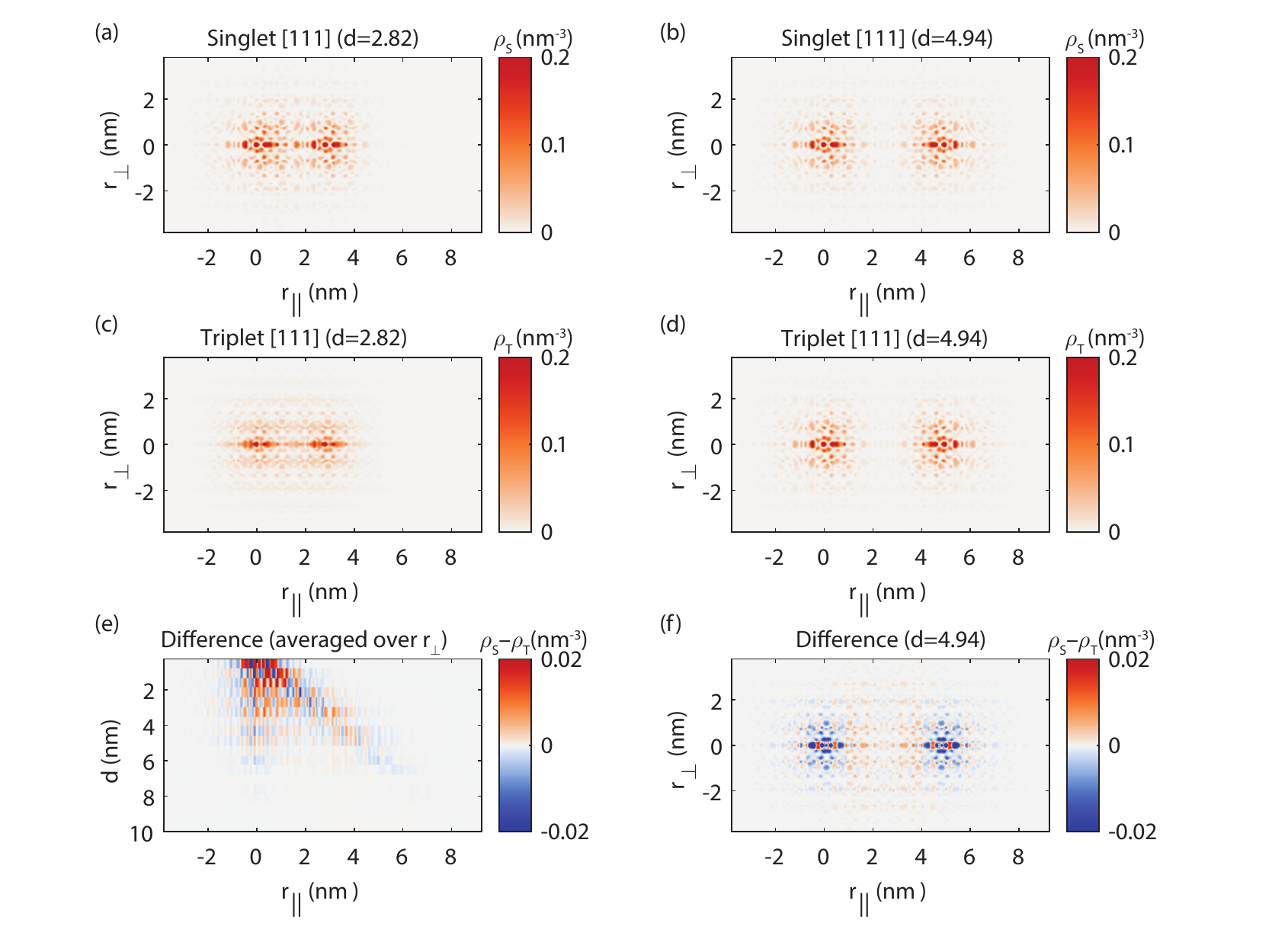}
	\caption{Electron density \req{Eq:ElectronDensity} calculations for one donor placed at the origin and the second moved along [111] with the slider. Shown is a cutplane defined by the axis joining the two donors ($r_{\parallel}$) and their cross-product with the z-axis $r_{\perp}$. The radio-button selects between the electron density of the lowest singlet ((a) and (b) offline), triplet  ((c) and (d) offline) or their difference  ((e) and (f) offline. (e) averages over the $r_{\perp}$-axis and iterates over all donor distances $d$ instead).}
	\label{Fig:8}
\end{figure}

Notably, in the large-distance regime we find that $J_{[100]}(d)>J_{[110]}(d)>J_{[111]}(d)$. This is due to the anisotropy of the envelope functions $F^{\mu}(\vec{r})$ (\req{Eq:EMTstate})\cite{voisin2020valley} as outlined in the following. Each $F^{\mu}(\vec{r})$ has an ellipsoidal shape squeezed along the valley orientation. The $+x$-valley combination of the $A_1$ ground state of the Bloch and envelope function  $|F^{+x}(\vec{r})\varphi_{+x}(\vec{r})|^2$ on the surface of a sphere with $r=\SI{10}{\nano\meter}$ around the donor is shown in \rfig{Fig:5}c. \rfig{Fig:5}d shows the full $A_1$ ground state. Each valley contributes a belt perpendicular to its axis. Coming from large distances, we have two $A_1$ states approaching each other and the exchange interaction strength will scale as the overlap of their wave functions. Along the [100] crystal axis ($x$-axis) two of these belts overlap and result in a high $J_{[100]}$. Along [110] (bisecting $x$ and $y$-axis) only the $z$-valley ellipsoids contribute fully giving an intermediate $J_{[110]}$, while going diagonally along [111] corresponds to the dark region in \rfig{Fig:5}d and a small $J_{[111]}$.

Another interesting feature is that the exchange interaction strength has a dent at roughly $d\approx\SI{4}{\nano\meter}$, especially for [100]. At this point the two lowest spin triplet states have an avoided crossing\cite{klymenko2014electronic}. This occurs when the spin triplet arising from an antibonding combination of $A_1$ orbitals (whose energy increases with decreasing distance) crosses the energy of the bonding combination of $T_2$ orbitals (whose energy decreases with decreasing distance). The consequence of this hierarchy inversion has been observed in an experiment with closely-spaced implanted donors \cite{dehollain2014single}.

The first row of \rfig{Fig:5}e shows the three lowest energy states $E^{S}_0(d)$ and $E^{T}_{0/1}(d)$ as a function of donor separation along the three crystal directions. 
The splitting between the $A_1$-like singlet (blue) and triplet (yellow) increases until the triplet meets the $A_1$-$T_2$- triplet (red). 
Along [110] and [111], we can additionally observe energy oscillations. 
The coupling between valley states depends on the symmetry of the system. The inter-donor valley-orbit coupling strength (\req{Eq:VOC}) crucially depends on the overlap of the Bloch functions on the individual donors. The plane wave component of the Bloch functions is not lattice periodic and results in interference patterns when the donors are moved relative to each other. This causes the energy oscillations for lower symmetry cubic crystal orientations, i.e. for all directions but the the $\braket{100}$.

A CROT operation relies on the magnetic drive of a single electron spin resonance line. On the one hand, $J$ needs to be large enough to allow the individual drive on a specific spectral line giving us a lower bound to $J$. A large $J$ will however induce deviations from the logical qubit basis given by the simple spin product states providing an upper boud to $J$. A high-fidelity CROT operation requires $\SI{0.1}{\mega\hertz}<J<\SI{10}{\mega\hertz}$ \cite{hill2007robust,kalra2014robust}, marked by the dashed lines in \rfig{Fig:5}a. Depending on the crystal direction this condition can be met starting from $d_{\rm min}=\SI{10.34}{\nano\meter}$ along [111] up to $d_{\rm max}=\SI{23.892}{\nano\meter}$ along [100]. Along [100] and [110], high-fidelity CROT gates are possible for donors placed over a range of $\SI{5}{\nano\meter}$. However, along [111] we find that the amplitude of the valley oscillations is of the order of the allowed exchange interaction strength range, and requires for precise donor placement. There is no distance range in which high-fidelity CROT gates are possible independent of the crystal orientation.

To analyse the valley contributions we define state projectors that have the valley structure as defined in \req{Eq:A1} to \req{Eq:E}. 
An $A_1$-like state of a donor at position $\vec{r}_b$ has components
\begin{align}
	\psi_{A_1}^{b}(\vec{r}_1) =\sum_{\mu}\sum_{a}\mathcal{N} A_1(\mu)\delta(\vec{r}_a-\vec{r}_b) e^{i\vec{k}_\mu\vec{r}_b} F^\mu_a(\vec{r}_1)\varphi_\mu(\vec{r}_1)
\end{align}
in analogy to \req{Eq:BasisExpansion} , where $a$ iterates the basis states in the $\mu$th valley. $\mathcal{N}$ is a normalization constant, $A_1(\mu)$ gives the relative valley amplitudes in \req{Eq:A1} and $\vec{r}_a$ is the center of the $a$th basis state. We compute the $A_1$-valley contribution
\begin{align}
	A_1(\Psi^{S/T}) = \sum_{b}\int|\braket{\psi_{A_1}^{b}(\vec{r}_1)|\Psi^{S/T}(\vec{r}_1,\vec{r}_2)}|^2 d^3\vec{r}_2,
	\label{Eq:ValleyContribution}
\end{align}
where $b$ sums over all donor positions, and similarly $T_2$ and $E$.
The second/third row of \rfig{Fig:5}e gives valley contributions of the lowest singlet/triplet. 
For large distances the electronic ground states of two separate donors are just $A_1$ like. Bringing them together breaks the $A_1$ symmetry. Doing so along [100], e.g. the $x$-axis,  will energetically favour $y$- and $z$-valley orbitals that are wide along this axis, and promote $E$ symmetry (\req{Eq:E}). At the point where the lowest triplet has an avoided crossing with the first excited triplet branch, it becomes roughly half $T_2$ like. Going along [111], favours no axis in particular, but lifts the directional symmetry along $x$, $y$ and $z$, introducing $T_2$ contributions (\req{Eq:T2}), while leaving $E$ states untouched. For short distances the triplet oscillates between large and small $T_2$ contributions when crossing with the first excited state. Along [110] the $z$-valleys have a strong overlap resulting in $E$ contributions, while the directional symmetry of the $x$- and $y$-valleys is broken giving small $T_2$ contributions to the ground state. Again the triplet $T_2$ contributions oscillate for short distances.

\begin{figure}[htb!]
	\includegraphics[width=1.0\columnwidth]{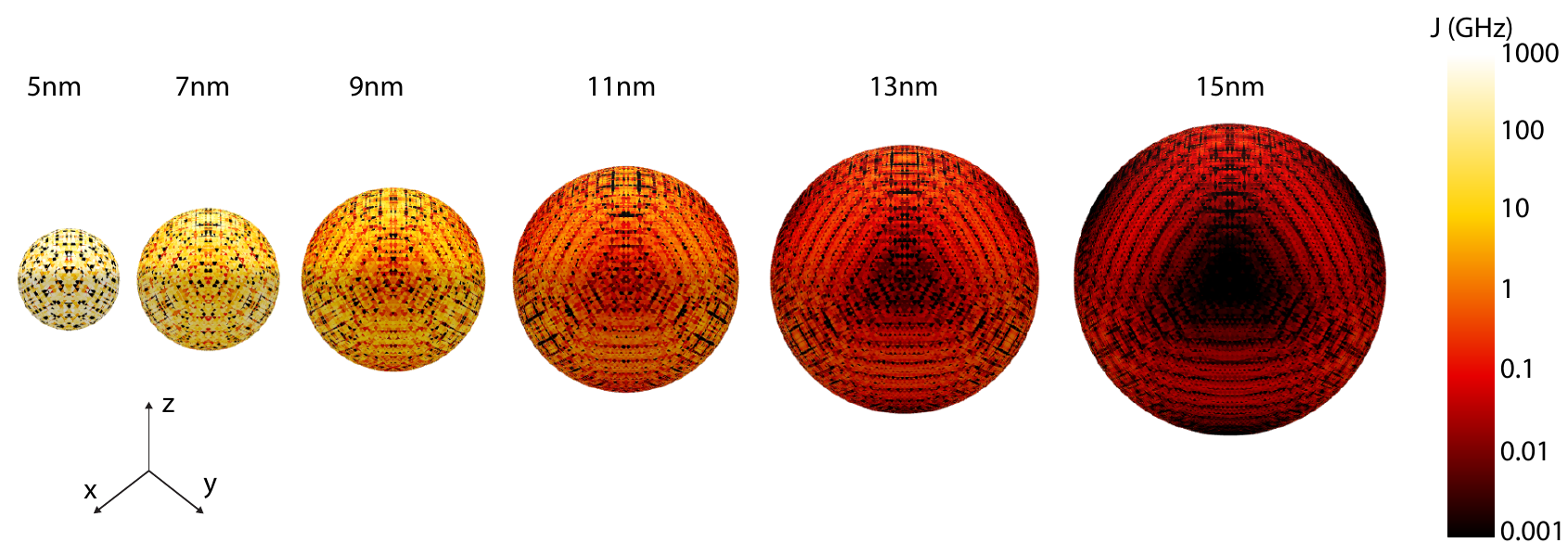}
	\caption{Exhaustive exchange interaction strength evaluation of two exchange-coupled donors in various spatial configurations. One donor is position in the center of the sphere, the second donor at the lattice site closest to each point on the surface. We show the resulting exchange interaction strength on a log scale. It is highly oscillatory and reproduces the anisotropy of the valley-orbit basis. }
	\label{Fig:9}
\end{figure}

\rfig{Fig:6}, \ref{Fig:7} and \ref{Fig:8} give the electron densities \req{Eq:ElectronDensity} for a donor placed at the origin and a second donor moved along the [100], [110] and [111] crystal axis. The cut-planes are defined by the axis joining the two donors and their cross-product with the z-axis. The slider moves the second donor, i.e. varies the inter-donor distance. The radio button changes between the electron density of the singlet, triplet or their difference. For large distances, they are predominately $A_1$-like and resemble the one-electron ground state\cite{gamble2015multivalley}. Additionally, the bonding singlet has more support in the region between the donors, than the anti-bonding triplet. Along [110] and [111], the amplitude of this discrepancy oscillates as a function of distances, due to the changing overlap of the Bloch functions at each donor position. We can observe the anisotropic structure of the wave functions on each donor that decay very fast along [111] and [110], while being wide along [100]. This leads to less support in between the donors reducing the singlet-triplet difference, i.e. the exchange interaction strength as discussed above. 
Additionally, at $d\approx\SI{4}{\nano\meter}$, the triplet state becomes more $T_2$-like which changes the lattice periodic structure of $\rho_T$. The bonding/anti-bonding structure of the $\rho_{S/T}$ ceases to exist leading to the dent in the exchange interaction, discussed above. Along [111], the second donor interchanges between inverted tetrahedral bonds, i.e. alternates between the shifted fcc sublattices. The electron density around this donor appears to periodically flip.

Finally, we perform exhaustive 3D iterations. In \rfig{Fig:9} we compute exchange interaction strength for one donor at the center of the shown spheres with $r=5-\SI{15}{\nano\meter}$ and the second on the lattice site closest to each point on the surface. 
We exploit the symmetry of the silicon crystal to reduce the number of unique calculations that are performed by a factor of 48.
We can nicely see all the features discussed above.

\section{Comparison to Literature} \label{Sec:Comparison}

\begin{figure}[htb!]
	\includegraphics[width=1.0\columnwidth]{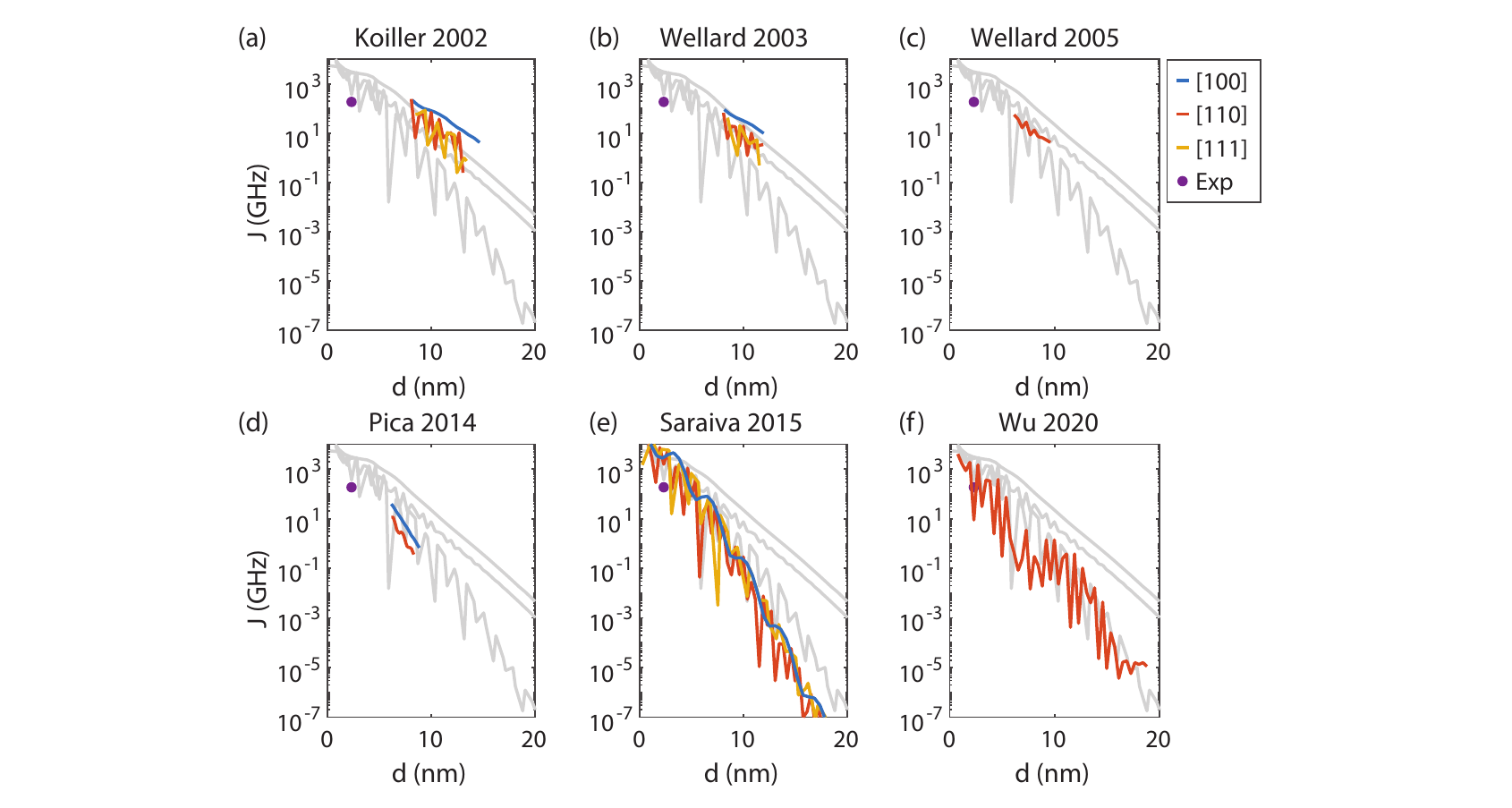}
	\caption{EMT exchange interaction strength calculations in literature as a function of distance along the [100], [110] and [111] crystal directions. The results presented in \rsec{Sec:ExchangeCoupledDonors} are shown as a grey shadow for easy comparison. Additionally, we include an experimental data point for $d=2.3\pm\SI{0.5}{\nano\meter}$ from \rref{gonzalez2014exchange} in purple (a), (b) \rref{koiller2001exchange} and \cite{wellard2003electron} used a Kohn-Luttinger EMT with a simple bulk-screened Coulomb potential. The exchange is calculated with the Heitler-London method. (c), (d) \rref{wellard2005donor} and \cite{pica2014exchange} also used Heitler-London, but included VOC and a CCC. (e) Gonzalez-Zalba \textit{et al.}\cite{gonzalez2014exchange} and Saraiva \textit{et al.}\cite{saraiva2015theory} implemented a Full CI code. Their EMT used an isotropic effective mass and orbitals, neglected VOC and used a CCC specific to the $A_1$ ground state. (f) \rref{wu2020excited} included the effective mass anisotropy, VOC and a CCC. Their GTO basis however is isotropic, the exchange is calculated via HF.}
	\label{Fig:10}
\end{figure}

\rfig{Fig:10} compares our exchange results to EMT implementations in the literature. The results of the cited works are shown in colour, while our results are shown as a grey shadow for comparison. Additionally, we include an experimental data point (purple) from \rref{gonzalez2014exchange}.

Early work by Koiller \textit{et al.}\cite{koiller2001exchange} and Wellard \textit{et al.}\cite{wellard2003electron} in \rfig{Fig:10}a and b used a Heitler-London approach, i.e. a single SD formed by the one-electron wave functions centered on each donor to model the singlet and triplet states. They approximated the impurity by a simple bulk-screened Coulomb potential and ignored VOC terms. 
They use a minimal basis with one anisotropic STO per valley per donor. The resulting exchange in \rfig{Fig:10}a is roughly an order of magnitude higher than the present work. Wellard \textit{et al.}\cite{wellard2003electron} are fairly consistent with our results. In both cases, the $J_{[110]}(d)$ and $J_{[111]}(d)$ levels have the same magnitude, but lie below the $J_{[100]}(d)$. In comparison, we found that $J_{[100]}(d)>J_{[110]}(d)>J_{[111]}(d)$.

Still using Heitler-London, Wellard \textit{et al.}\cite{wellard2005donor} and Pica \textit{et al.}\cite{pica2014exchange} included isotropic CCCs from \rref{pantelides1974theory} and \cite{ning1971multivalley} respectively. They included VOC terms, but still used the same minimal basis set.
Their EMT failed to precisely reproduce the excited valley state energies. Again, the results by Wellard \textit{et al.}\cite{wellard2005donor} in \rfig{Fig:10}c are consistent with ours, while the exchange in \rfig{Fig:10}d is over an order of magnitude smaller.

Gonzalez-Zalba \textit{et al.}\cite{gonzalez2014exchange} and Saraiva \textit{et al.}\cite{saraiva2015theory} implemented a Full CI as the current work. However, they used a hydrogenic model, i.e. an isotropic basis and effective mass. As a result, the exchange in \rfig{Fig:10}e has similar amplitudes along all directions, i.e.  $J_{[100]}(d)\approx J_{[110]}(d)\approx J_{[111]}(d)$. Their basis was made from one STO-3G orbital per valley per donor and they neglected VOC. Additonally, they applied a CCC that interpolates between the bulk-screened and vacuum case to the donor ground state and assumed the excited states to be degenerate. In this model, the exchange decays quickly for all directions comparable to our [111] results. Notably, the exchange along [100] is not flat there, but exhibits a step like decay.

Wu \textit{et al.}\cite{wu2020excited} used an EMT including VOC and a CCC in the style of Saraiva \textit{et al.}\cite{saraiva2015theory}. While their basis consisted of five isotropic GTOs per valley per donor, their kinetic energy operator included the effective mass anisotropy. Again, these approximations lacked the precision to produce the excited valley state energies. To obtain the results in \rfig{Fig:10}f, they used the HF method described in \rsec{Sec:HartreeFock}. Their [110] results are comparable to our [111] calculations. Note that the exchange seems to saturate at large distances.

\section{Electric Tunability and SWAP Gates}
\label{Sec:Tunability}
\begin{figure}[htb!]
	\includegraphics[width=1.0\columnwidth]{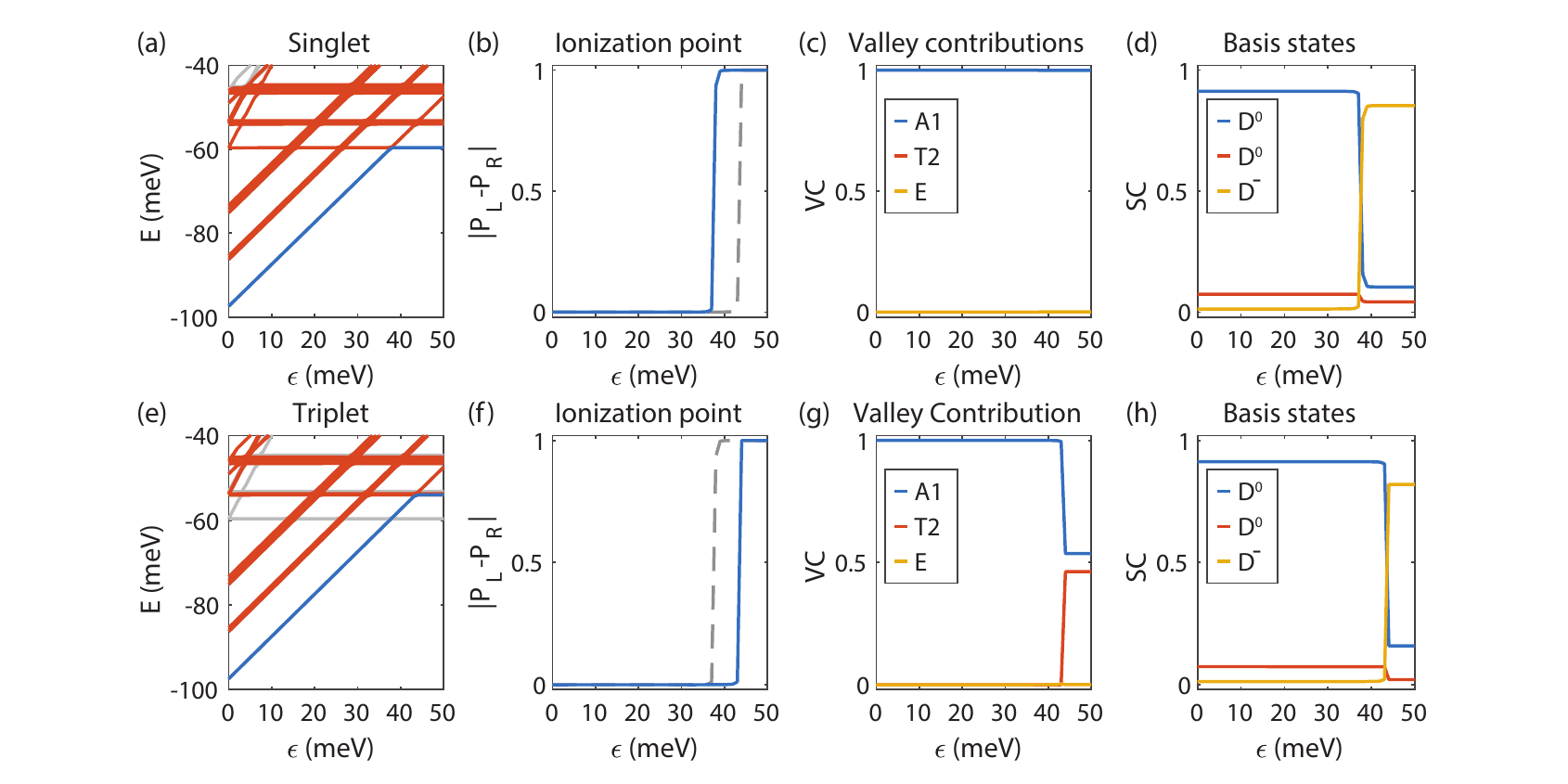}
	\caption{Analysis of the electrical tunability of two phosphorus donors positioned at distance $d=\SI{20.09}{\nano\meter}$ along the [100] crystal axis. An electric field $E_x$ is applied along the axis joining the donors and we vary the detuning $\varepsilon=E_x d$. The first/second row looks at the singlet/triplet states. (a), (e) The energies of the singlet (a) or triplet (e) subspace as a function of detuning. In each panel, the energies of the spin other subspace are shown in grey. The lowest energy states analyzed in the following panels are marked in blue. (b), (f) The absolute difference in probability of finding an electron on the left or right donor. (c), (g) The valley contributions computed via \req{Eq:ValleyContribution}. Upon ionization the triplet becomes $A_1$-$T_2$-like. (d), (h) Contributions of the different basis states computed via \req{Eq:StateContribution}.}
	\label{Fig:11}
\end{figure}
\begin{figure}[htb!]
	\includegraphics[width=1.0\columnwidth]{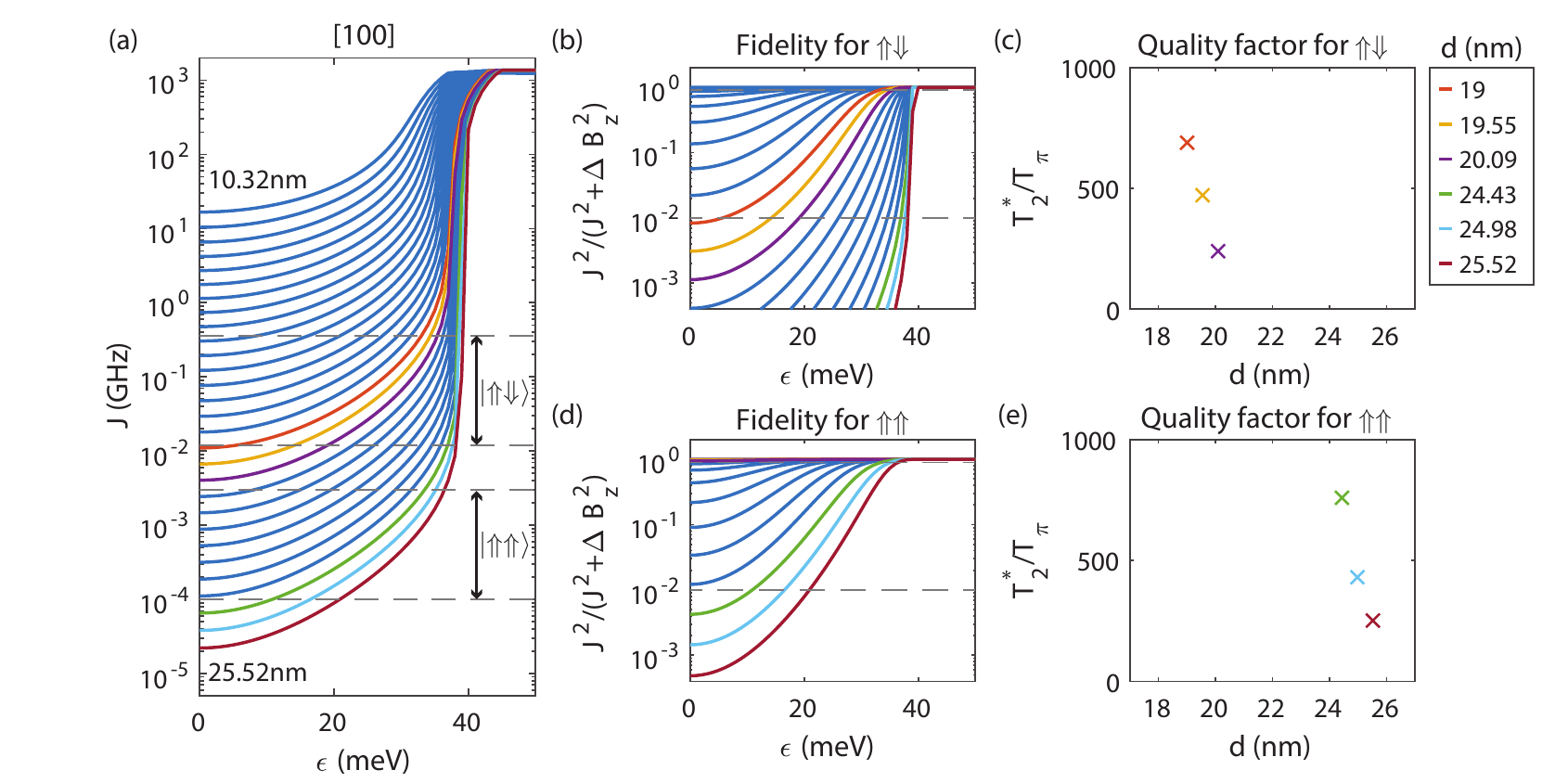}
	\caption{Viability analysis of a SWAP gate performed on two tunable phosphorus donors positioned along [100]. (a) Exchange interaction strength as a function of detuning $\varepsilon$ for distances between \SI{10}{\nano\meter} and \SI{25.5}{\nano\meter}. $J$ increases with field until the system is ionized and $J$ saturates. The amplitude of the SWAP operations are given by $F=J^2/(J^2+\Delta B_{z}^2)$ for anitparallel nuclear spins $\Uparrow\Downarrow$ with $\Delta B_{z}=\SI{120}{\mega\hertz}$ in (b) and parallel nuclear spins $\Uparrow\Uparrow$ with $\Delta B_{z}=\SI{1}{\mega\hertz}$ in (d). For high-fidelity gates, we need to tune $F$ from $0.01$ (lower dashed line) to $0.9$ (upper dashed line). The corresponding exchange tuning ranges are indicated in (a). (c)/(e) For viable distances, we evaluate the quality factor in \req{Eq:Quality} (number of coherent oscillations) at the $F=0.9$ points.}
	\label{Fig:12}
\end{figure}
\begin{figure}[htb!]
	\includegraphics[width=1.0\columnwidth]{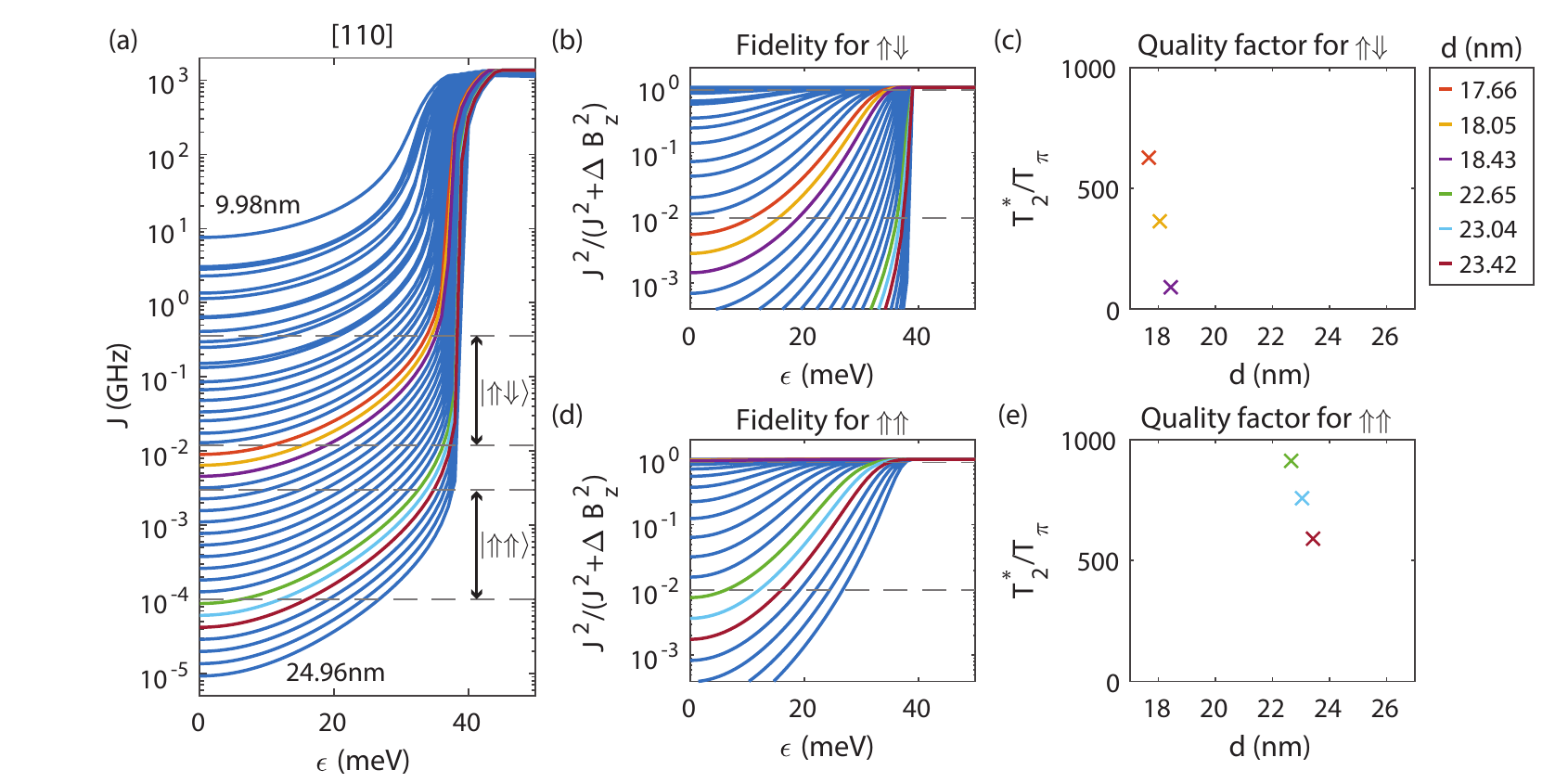}
	\caption{Viability analysis of a SWAP gate performed on two tunable phosphorus donors positioned along [110]. (a) Exchange interaction strength as a function of detuning $\varepsilon$ for distances between \SI{10}{\nano\meter} and \SI{25}{\nano\meter}. $J$ increases with field until the system is ionized and $J$ saturates. The amplitude of the SWAP operations are given by $F=J^2/(J^2+\Delta B_{z}^2)$ for $\Delta B_{z}=\SI{120}{\mega\hertz}$ in (b) and $\Delta B_{z}=\SI{1}{\mega\hertz}$ in (d). For high-fidelity gates, we need to tune $F$ from $0.01$ (lower dashed line) to $0.9$ (upper dashed line). For viable distances, we evaluate the corresponding quality factor \req{Eq:Quality} (number of coherent oscillations) at the $F=0.9$ points in (c) and (e).}
	\label{Fig:13}
\end{figure}

In the previous section we discussed a pair of exchange-coupled phosphorus donors as a function of their distance $d$, in the absence of electric fields. Now we investigate the electric field dependence of $J$, with the goal of understanding under which conditions it becomes possible tune the exchange interaction to perform two-qubit SWAP gates \cite{kane1998silicon,hollenberg2006two,hill2007robust}. We add an electric field potential to our EMT Hamiltonian in \req{Eq:EMTHamiltonian}
\begin{align}
	U_E(\vec{r})= \vec{E}\cdot\vec{r},
\end{align}
where $\vec{E}$ is the electric field vector. We neglect VOC terms for $U_E$, due to their fast oscillatory behaviour.
In the following, we will apply a field along the axis joining the donors and define the detuning $\epsilon=|\vec{E}|d$, where $d$ is the donor separation.
We tested the effect of polarization when $p$-orbitals ($n_i=1$) are included. We found that polarization is only a minor effect and to first order the system can be explained in a molecular orbital approximation containing covalent and ionic contributions, i.e. $D^{0}$ and $D^{-}$ orbitals.

\rfig{Fig:11}a and e show exemples of singlet/triplet energies of two donors positioned along [100] at $d=\SI{20.09}{\nano\meter}$ as a function of detuning. States with one electron on each donor are sensitive to a detuning and have a slope, while ionized states are approximately flat.
With increasing detuning, the lowest singlet and triplet states (blue) increase in energy until the ionized state becomes energetically favourable. The singlet allows closed-shell configurations, i.e. $i=j$ in \req{Eq:SingletSAC}. As a result, the ionized singlet is just $A_1$-$A_1$-like.
The triplets in \rfig{Fig:11}e do not allow such solutions and the lowest ionized state is $A_1$-$T_2$-like and we are missing the equivalent to the $A_1$-$A_1$-like singlet shown in grey.
As a result, we need to invest more energy to ionize the triplet and the ionization point in \rfig{Fig:9}f is shifted compared to the singlet in \rfig{Fig:11}b.
The valley contributions (\req{Eq:ValleyContribution}) in \rfig{Fig:9}c and g tell the same story.
For zero detuning, the singlet and triplet are both $A_1$ like.
At the ionization point, the triplet becomes $A_1$-$T_2$-like.
\rfig{Fig:11}d and h give the state contributions (\req{Eq:StateContribution}) as a function of detuning. At the ionization point the states occupy $D^{-}$-like basis functions.

We now analyze the viability of tunable exchange-coupled donors to perform SWAP gates. The SWAP oscillations of two spins prepared in $\ket{\uparrow\downarrow}$ are described by the Rabi formula\cite{rabi1937space}
\begin{align}
	P_{\ket{\uparrow\downarrow}\rightarrow\ket{\downarrow\uparrow}}(t)=\frac{J^2}{J^2+\Delta B_{z}^2}\sin^2\left(\sqrt{J^2+\Delta B_{z}^2}\frac{t}{2}\right),
	\label{Eq:Rabi}
\end{align}
where $J$ is the exchange interaction strength and $\Delta B_{z}$ the longitudinal magnetic field gradient across the two electrons. Tuning $J$ relative to $\Delta B_{z}$ allows us to tune the prefactor in \req{Eq:Rabi} and switch the SWAP oscillations on or off. 

In a donor system, the electron and the nuclear spin are coupled by a hyperfine interaction $A\approx 120$~MHz, which in a gated nanoscale device can vary from one donor to the next by an order $\Delta A \approx 1$~MHz. This allows the introduction of an intrinsic $\Delta B_{z}$, switchable by controlling the state of the two nuclear spins \cite{kalra2014robust,mkadzik2020conditional}.

If the nuclei are parallel, a small $\Delta B_{z}=\Delta A\approx \SI{1}{\mega\hertz}$ is created by the difference in the hyperfine interaction between the two donors. A much larger $\Delta B_{z}=\bar{A}\approx \SI{120}{\mega\hertz}$ can be obtained by preparing the nuclei in an antiparallel state.

\rfig{Fig:12}a shows the exchange interaction strength (\req{Eq:Exchange}) as a function of detuning for donors positioned along [100] at distances between \SI{10}{\nano\meter} and \SI{25.5}{\nano\meter}. 
Here, the smallest distance belongs to the curve with the highest $J$.
In the unperturbed system, the exchange interaction strength is the weakest for a given distance and increases as a function of $\epsilon$.
After the singlet ionization point (see \rfig{Fig:9}), the exchange interaction has a steep increase and finally saturates after the triplet ionization.
\rfig{Fig:12}b and d plots the prefactor in \req{Eq:Rabi} $F=J^2/(J^2+\Delta B_{z}^2)$ for $\Delta B_{z}=\SI{120}{\mega\hertz}$ and $\Delta B_{z}=\SI{1}{\mega\hertz}$.
For a high fidelity SWAP gate we demand to tune this from $F=0.01$ to $0.9$ indicated by the dashed lines.
To perform the gate, we prepare the system anywhere below the $F=0.01$ line, tune it to the $F=0.9$ line, then let it evolve for $T_{\pi}=\pi/\sqrt{J^2+\Delta B_{z}^2}$ and finally pulse it back below the $F=0.01$ line.
The corresponding exchange interaction tuning ranges are indicated in \rfig{Fig:12}a.
To estimate the quality of the SWAP oscillations at $F=0.9$, we assume quasistatic noise, where the decay time is given by\cite{dial2013charge}
\begin{align}
	T_{2}^{*}=\frac{1}{\sqrt{2}\pi(dJ/d\epsilon)\epsilon_{rms}},
\end{align}
where we take $\epsilon_{rms}=\SI{24e-3}{\giga\hertz\per\nano\meter} \times d$ as our noise amplitude\cite{freeman2016comparison,thorgrimsson2017extending,harvey2017coherent}.
We then define the quality factor as the number of SWAP oscillations we can perform in $T_{2}^{*}$
\begin{align}
	Q=\frac{T_{2}^{*}}{T_{\pi}}=\sqrt{\frac{J^2+\Delta B_{z}^2}{2(dJ/d\epsilon)^2\epsilon_{rms}^2}}.
	\label{Eq:Quality}
\end{align}
\rfig{Fig:12}c, e give the quality factor $Q$ at the $F=0.9$ point in \rfig{Fig:12}b, d for viable distances.
The quality factors are typically larger than 100 and on the same order of magnitude for parallel and antiparallel nuclear spin configurations. Picking  distances larger than the minimal requirement, will result in a steeper slope of $J$ at the driving point in \rfig{Fig:12}a. This increases the sensitivity to charge noise, which leads to a shorter $T_{2}^{*}$ and the decrease in quality factor with distance in \rfig{Fig:12}c,e.

\rfig{Fig:13} produces the same graphs for two tunable donors positioned along [110]. Here, the same range of exchange coupling is obtained at shorter distances due to the effective mass anisotropy (see \rsec{Sec:ExchangeCoupledDonors}). Additionally, we can see the valley oscillations, which cause the irregular decrease of $J$ with distance. We obtain quality factors of the same order of magnitude as for the [100] crystal orientation. In this configuration parallel nuclear spins yield slightly better results. However, the using such small values of exchange interaction would require slow $\gg \SI{1}{\micro\second}$ pulses to avoid spectral broadening beyond the value of $J$, whereas antiparallel spins allow much faster gate operations, on the order of $\SI{100}{\nano\second}$.

\section{Conclusion}
\label{Sec:Conclusion}
We presented Full Configuration Interaction simulations of exchange-coupled donors using a state-of-the-art implementation of multi-valley effective mass theory. We tested two basis sets and found the set summarized in \rtab{Tab:SmallBasis} to be sufficient for donor separations $d\gtrapprox\SI{5}{\nano\meter}$. This allowed us to perform 3D iterations including $8500$ donor configurations. We studied in detail two donors positioned along [100], [110] and [111]. We analysed the exchange interaction, the valley configurations and visualized the evolution of electron density as a function of distance. We found that a high-fidelity CROT gates are possible over a range of distances between \SI{10}{\nano\meter} and \SI{24}{\nano\meter} depending on the crystal direction. Along [100] and [110] the donor placement may straggle by \SI{5}{\nano\meter}. Finally, we explored the electric tunability of exchange-coupled donors. High-fidelity SWAP gates require the system to be pulsed close to the ionization point. However, using a simple noise model we predict that high-quality SWAPs are possible.
This work shows how efficient modelling can inform the design of two-qubit devices.
\newpage
\appendix
\section{STO-nG fits}
\label{App:STOnG}
\begin{figure}[h]
	\centering
	\includegraphics[width=0.5\columnwidth]{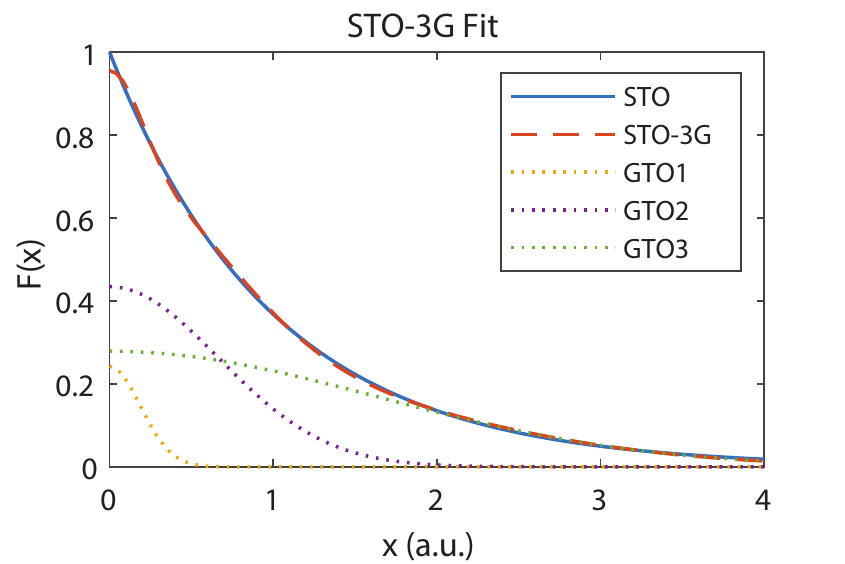}
	\caption{Visulatization of the STO-$n$G fits (see \req{Eq:STOnG}).}
	\label{Fig:App1}
\end{figure}
\FloatBarrier
\section{Carbon footprint}
\label{App:Carbon}
\begin{table}[h]
	\caption{Carbon emissions involved in all calculations run in this project. We included all calculations, even if the results are not shown in this publication. Estimations follow the standardised reporting table by Scientific CO$_2$nduct\cite{co2nduct}. }
	\centering
	\label{Tab:CO2Footprint}
	\begin{tabular}{ll}
		\hline
		\textbf{Numerical Simulations}\\
		\hline
		Total Kernel Hours (h) & 181130 \\
		Thermal Design Power Per Kernel (W) & 5.75 \\
		Total Energy Consumption Simulations (kWh) & 1041.5  \\
		Average Emission Of CO$_2$ In New South Wales\cite{dee2020national} (kg/kWh) &	0.81  \\
		Total CO$_2$-Emission For Numerical Simulations (kg) & 843.6  \\
		Were The Emissions Offset? & \textbf{No}  \\
		\hline
		\textbf{Transport}\\
		\hline
		Total CO$_2$-Emission for Transport (kg) & 0\\
		Were The Emissions Offset? & n/a  \\
		\hline
		Total CO$_2$-Emission (kg) & 843.6\\
		\hline
		\hline
	\end{tabular}
\end{table}
\FloatBarrier
\acknowledgments
The research at UNSW was funded by the Australian Research Council Centre of Excellence for Quantum Computation and Communication Technology
(Grant No. CE170100012) and the US Army Research Office (Contract No. W911NF-17-1-0200). This research/project was undertaken with the assistance of resources and services from the National Computational Infrastructure (NCI), which is supported by the Australian Government.
The research at Sandia National Laboratories was supported by the Laboratory Directed Research and Development program, Project 213048. Sandia National Laboratories is a multi-missions laboratory managed and operated by National Technology and Engineering Solutions of Sandia, LLC, a wholly owned subsidiary of Honeywell International Inc., for the National Nuclear Security Administration of the US Department of Energy under contract DE-NA0003525.
The views and conclusions contained in this document are those of the authors and should not be interpreted as representing the official policies, either expressed or implied, of the ARO or the US Government. The US Government is authorized to reproduce and distribute reprints for government purposes notwithstanding any copyright notation herein.

\bibliography{main}
\end{document}